\documentclass[english]{article}
\usepackage[T1]{fontenc}
\usepackage[latin9]{inputenc}
\usepackage{geometry}
\usepackage{float}
\usepackage{url}
\usepackage{amsmath}
\usepackage{graphicx}

\makeatletter
\usepackage[breaklinks,colorlinks=true,linkcolor=black, citecolor=blue, urlcolor=blue]{hyperref}
\usepackage{tocbibind}

\makeatother

\usepackage{babel}
\begin{document}
\title{Leveraged ETF Investing}
\author{Tal Miller}
\maketitle
\begin{abstract}
It is common knowledge that leverage can increase the potential returns
of an investment, at the expense of increased risk. For a passive
investor in the stock market, leverage can be achieved using margin
debt or leveraged-ETFs. We perform bootstrapped Monte-Carlo simulations
of leveraged (and unleveraged) mixed portfolios of stocks and bonds,
based on past stock market data, and show that leverage can amplify
the potential returns, without significantly increasing the risk for
long-term investors.
\end{abstract}
\newpage{}

\tableofcontents{}

\newpage{}

\section{Introduction}

Leverage (borrowing to invest) is a way to increase potential returns
for an investment, at the expense of increased risk. For a passive
investor in the stock market, this can be achieved by taking margin
loan from the brokerage, or buying leveraged exchange traded funds
(LETFs) \cite{FelixLETF}. LETFs (that amplify the daily returns of
their underlying index) are usually not recommended as long term investments
due their decay during fluctuations (even when the index ``fluctuates''
around a constant value the LETF loses) \cite{investopediaLETFs,etfcomLETFs,RichMoose}.
Margin leverage is less sensitive to daily fluctuations which makes
it interesting to compare both methods. An additional leverage strategy
that can be employed is using stock options (e.g. constantly buying
call options), but the author does not currently understand them enough
to model.

The stock market is chaotic and the price swings are almost uncorrelated
day to day. Predicting the future is impossible, but over the long
term the market goes up, and this has been consistent for hundreds
of years. The daily price change (in percents) $\Delta P$ can be
considered a random variable whose probability distribution is skewed
upwards slightly. Given that we have no better knowledge about the
future except the past, the best we can do is to assume that $\Delta P$
in the future is distributed the same as in past. Generating synthetic
realizations of the future by drawing from this distribution is called
a Monte-Carlo (MC) simulation, which is classically done by assuming
some analytic probability distribution for the price changes. The
result is a probability distribution for the yield of a portfolio
after some investment period (say, 10 years), for which we can calculate
risk and reward metrics. Note that for the simulation to be realistic,
it need to capture the statistical correlations between different
asset classes. One can also draw from the $\Delta P$ distribution
by picking price changes from random days in the actual past data,
this is as realistic as possible and captures the correlations between
different asset classes. We shall call this bootstrapped Monte-Carlo
(BMC) and use it in this work \cite{KositskyMedium}.

A mixed stocks/bonds portfolio is beneficial in reducing the fluctuations
over time, due to the fact that stocks and bonds are different asset
classes that are somewhat anti-correlated \cite{WikiMPT,WikiMarkowitz,Markowitz1952}.
Using backtesting and the Monte-Carlo method it was shown that leveraged
stocks/bonds portfolios (risk parity) can boost the risk-adjusted
returns \cite{Hedgefundie,DiLellio2018,Ayres2013,WealthfrontRiskParity}.
It is important to note that the future is not the same as the past
and Monte-Carlo is limited that way \cite{MindfullyInvest}, but if
we want to make the least amount of assumptions about the future,
this is the best method we have for quantifying future risk/reward
and comparing between different strategies.

In this work we perform BMC simulations ourselves and quantify the
performance of leveraged portfolios, using both LETFs and margin.
In addition, we do the calculation for both tax-free (IRA account
in Israel) and taxable accounts. We show how applying leverage on
a mixed stocks/bonds portfolio can greatly amplify the yield without
significantly increasing the risk.

\textbf{Many thanks to Roman Kositsky and Itay Katzir for significant
contributions.}

\section{Data \label{sec:Data}}

In this chapter we present the raw data we use in this work.

\subsection{Stocks}

For stocks we use the SP500 and NDX100 indices. The autor is mostly
interested in the NDX100 (due to its technology bias) which began
in 1985, so we will not look at earlier years for both indices for
simplicity. The indices data is from Yahoo \cite{Yahoo}, but SP500-TR
(total return, including dividends) data is only from mid-1988, so
we confine all the simulation to use data beginning at 1.1.1989. The
maximal date of the data is 30.9.20, when the extraction was performed.

The SP500 and SP500-TR data:

\begin{figure}[H]
\begin{centering}
\includegraphics[scale=0.5]{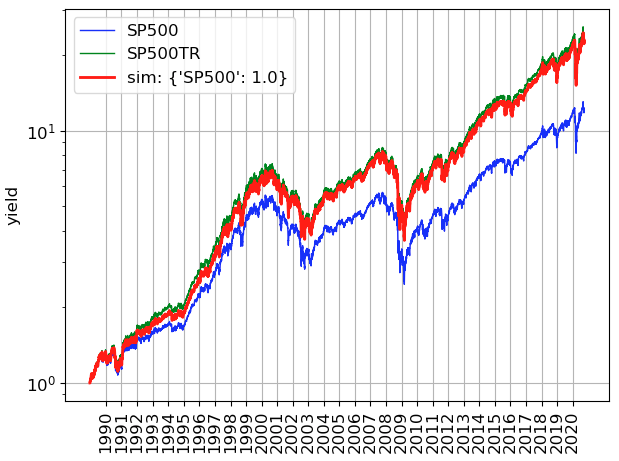}
\par\end{centering}
\caption{Data for the SP500 index (price and total return) for the 1989-2020
period. A tax-free simulation is plotted in red.}
\end{figure}

We also show a simulation in red (discussed in chapter \ref{sec:Portfolio-Evolution-Algorithms})
which follows the SP500 price data and adds (yearly) dividends of
2\% (for simplicity we calibrated a constant number for the whole
tested period, and it fits pretty good for our purpose).

The NDX100 data is from Yahoo \cite{Yahoo}, but there is no corresponding
NDX100-TR data. The best we found was \cite{NDX100TR}, which begins
in 1999, so prior to that we artificically created NDX100-TR as NDX100
+ assumed 0.7\% dividends:

\begin{figure}[H]
\begin{centering}
\includegraphics[scale=0.5]{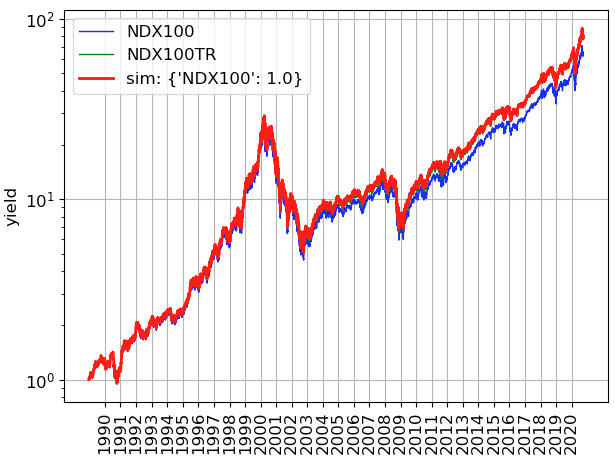}
\par\end{centering}
\caption{Data for the NDX100 index (price and total return) for the 1989-2020
period. A tax-free simulation is plotted in red.}
\end{figure}

Again, a simulation is shown in red that perfectly fits the TR data
(green, hidden behind the red). Before 1999 the simulation and data
coincide by construction, and after 1999 the constant 0.7\% dividend
works well. The NDX100 in general has far less dividends so the effect
is small.

\subsection{Bonds}

For bonds we use the long term treasury ETF VUSTX, since its inception
is in 1986, prior to 1989. The data is again from Yahoo \cite{Yahoo},
where the price is the ``Close'' column and the total return is
the ``Adj Close'' column:

\begin{figure}[H]
\begin{centering}
\includegraphics[scale=0.5]{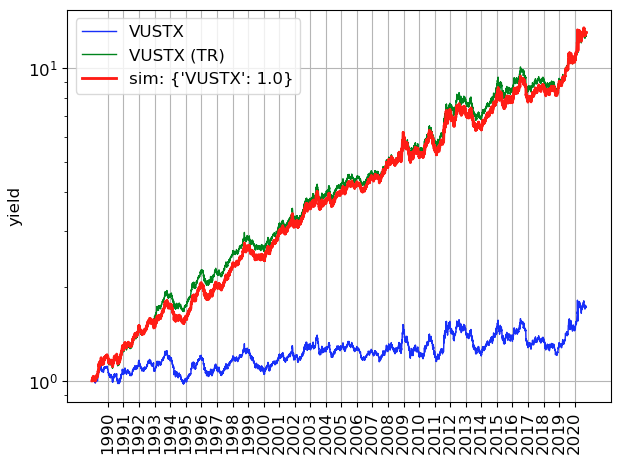}
\par\end{centering}
\caption{Data for the long term treasury bond ETF VUSTX (price and total return)
for the 1989-2020 period. A tax-free simulation is plotted in red.}
\end{figure}

First, we can see the significant contribution of the dividends to
the total return of holding bonds. The simulation is in red, but the
dividend model used is slightly more complicated than a constant percentage
we used before. What we use is actually $5\%+0.5\times\left[\text{1 month LIBOR rate}\right]$,
where the LIBOR rate is time dependent and taken from \cite{LIBOR}:

\begin{figure}[H]
\begin{centering}
\includegraphics[scale=0.5]{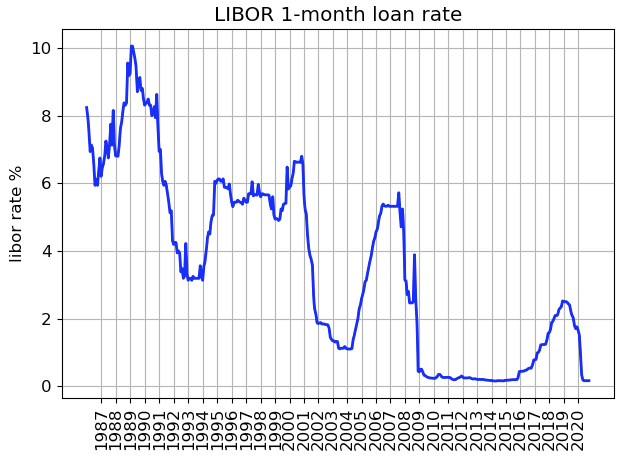}
\par\end{centering}
\caption{Data for the 1-month LIBOR rate for the 1986-2020 period.}
\end{figure}

We can see it changed significantly over the years, which is why a
constant \% fit does not work well.

\section{Portfolio Evolution Algorithms \label{sec:Portfolio-Evolution-Algorithms}}

In this chapter we describe the algorithms we use in the portfolio
evolution simulations, as implemented in \cite{mycode}. To validate
the simulation, we test it against real data.

\subsection{Buying papers}

We model the purchases of shares as ``papers'' (for lack of a better
name) of a specific stock type (the name of the ETF) which can have
an arbitrary value, and not an integer number. This should not matter
when the portfolio value is much larger than the cost of single share.
Each paper's value at purchase is logged since it matters for capital
gains tax.

Buying and selling stocks costs transaction fee from the brokerage.
We take this into account by applying a constant $0.1\%$ fee of the
value of each transaction, even though in real brokerages such as
Interactive Brokers the fee scheme is not given in a percentage but
in a more complicated form. We do not deal with this here and resort
to a more simplistic approach, which in fact does not affect the main
conclusions in a meaningful way.

The portfolio will be defined by the relative amounts of stocks and
bonds. The ideal (or target) portfolio fractions are defined by $f_{i}^{ideal}$,
so they satisfy $\sum_{i}f_{i}^{ideal}=1$. We start off the portfolio
by splitting the initial investment and buying $f_{i}^{ideal}$ from
each stock type.

\subsection{Price evolution}

We simulate how each paper's value evolves from day to day using closing
prices data.

\subsubsection{ETF evolution}

ETFs follow some index that change in value (in percents) by $dp_{index}$.
ETFs also have expenses, called expense ratio $ER$, which is some
percentage per year. The fee is charged daily, and since there are
252 trading days a year, the paper value decreases daily by $ER/252$
percents. Overall, the daily price change of the ETF is:

\begin{equation}
dp=dp_{index}-\frac{ER}{252}
\end{equation}

For the SP500 index we use the VOO ETF that has $ER=0.03\%$ and for
NDX100 we use QQQ ETF that has $ER=0.2\%$, significantly higher compared
to VOO. We do not have an ``index'' for bonds so we are using the
VUSTX price itself as the index, which is an ETF with $ER=0.05\%$. 

\subsubsection{Dividends}

The dividend rate we use for the stocks and bonds were defined in
chapter \ref{sec:Data}. For simplicity we add them to the cash balance
of the account on a daily basis. So if the paper value is $V$ and
yearly dividend rate (in percents) is $d$, the cash added daily due
to this paper is $Vd/\left(100\cdot252\right)$. 

Examples for how well the simulation of expense ratio and dividends
compare to data were shown in chapter \ref{sec:Data}.

\subsubsection{Cash reinvestment}

Once a month we reinvest any available cash to buy additional papers,
this cash can come either from dividends or from additional periodic
investements. We do not simulate the latter in this work, but the
option exists in the code \cite{mycode}.

As the portfolio evolves the portoflio fractions change dynamically
from $f_{i}^{ideal}$ to $f_{i}$. As we reinvest we can try and rebalance
by buying papers in different quantities. However, it is most likely
that the avaiable cash is not enough to properly rebalance (when the
portoflio becomes large enough significant selling will be required
to rebalance). We leave the rebalancing step to section \ref{subsec:Rebalancing},
and simply split the cash by $f_{i}^{ideal}$ for each stock type.

\subsubsection{Leveraged-ETF evolution}

The leveraged-ETFs (LETFs) we discuss in this work amplify the daily
returns of their underlying index. This comes at a price, the expense
ratio is around $\sim1\%$ (around 30 times higher than of VOO, and
5 times higher than of QQQ), and in addition the loans they take cost
as well. In \cite{Hedgefundie} it was estimated that we should use
the LIBOR rate as an additional expense for every 100\% of leverage.
Meaning, if we define the (1-month) LIBOR rate (in percents, yearly)
by $LR$, and the ETF leverage factor by $L_{ETF}$, then the additional
expense rate is $LR\left(L_{ETF}-1\right)$.

Another important note is that the LETFs are synthetic products that
for the most part do not own the actual stocks and track the total
return (TR) of the underlying index, meaning the dividends contribution
is already priced in. In practice, they do own some stocks and so
pay small amount of dividends, but we will neglect this in the simulation
and assume no dividends are given.

Overall, the daily price change of the LETF is:

\begin{equation}
dp=dp_{index,TR}L_{ETF}-\frac{ER}{252}-\frac{LR\left(L_{ETF}-1\right)}{252}
\end{equation}

The leveraged ETFs we simulate are:
\begin{itemize}
\item For the SP500 index the 2X LETF is SSO, and the 3X LETF is UPRO (the
underyling index of these LETFs is SP500-TR).
\item For the NDX100 index the 2X LETF is QLD, and the 3X LETF is TQQQ (the
undryling index of these LETFs is NDX100-TR).
\item For the bonds, an existing alternative to VUSTX that has LETFs that
leverage it is TLT, whose 2X LETF is UBT, and 3X LETF is TMF. We use
the VUSTX ETF since it has the longest historic data, and call its
fictitious LETFs by VUSTX2 and VUSTX3 (whose underlying index is VUSTX-TR). 
\end{itemize}
The reason we need to simulate the LETFs rather than simply using
existing data is because they are relatively new, and we want to ``extend''
them backwards to 1989. Also, when we perform Monte-Carlo simulations
in chapter \ref{sec:Monte-Carlo-simulations} we are going to generate
synthetic realizations of stock histories for which we have to simulate
how the LETF would evolve.

\paragraph{Examples}

We now validate our simulation against real LETF data to show it works
well.

Simulation of TMF (assuming TLT-TR as the underlying index) against
data, starting at 2011 (approximately when TMF started):

\begin{figure}[H]
\begin{centering}
\includegraphics[scale=0.7]{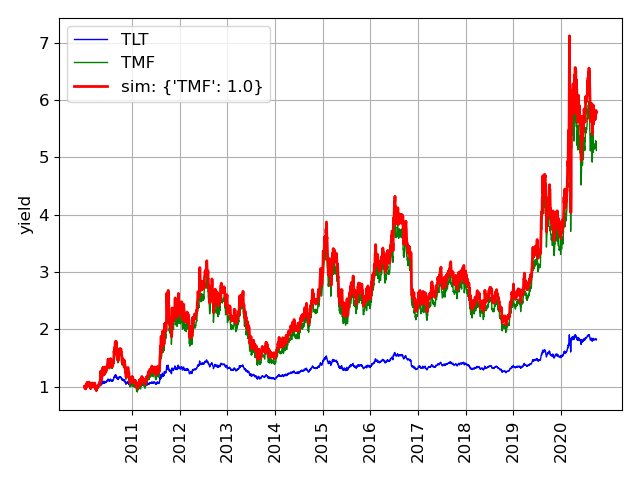}
\par\end{centering}
\caption{Data for the TLT and TMF ETFs for the 2010-2020 period. A tax-free
simulation of TMF (based on total return of TLT) is plotted in red.}
\end{figure}

Simulation of UPRO against data, starting at 2010 (approximately when
UPRO started):

\begin{figure}[H]
\begin{centering}
\includegraphics[scale=0.7]{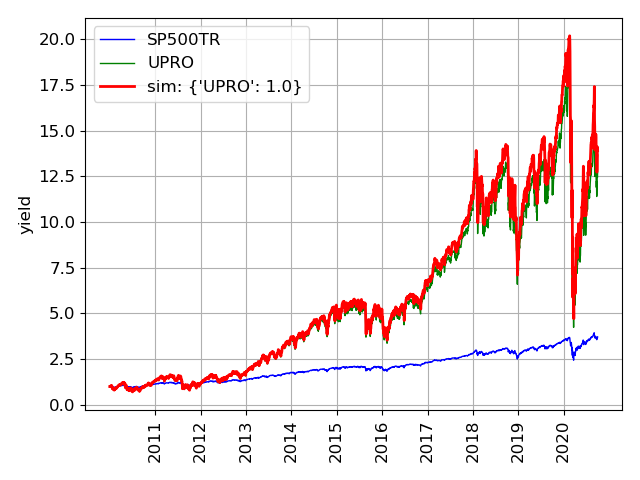}
\par\end{centering}
\caption{Data for the SP500-TR index and UPRO ETF for the 2010-2020 period.
A tax-free simulation of UPRO (based on SP500-TR) is plotted in red.}
\end{figure}

Simulation of TQQQ against data, starting at 2011 (approximately when
TQQQ started):

\begin{figure}[H]
\begin{centering}
\includegraphics[scale=0.7]{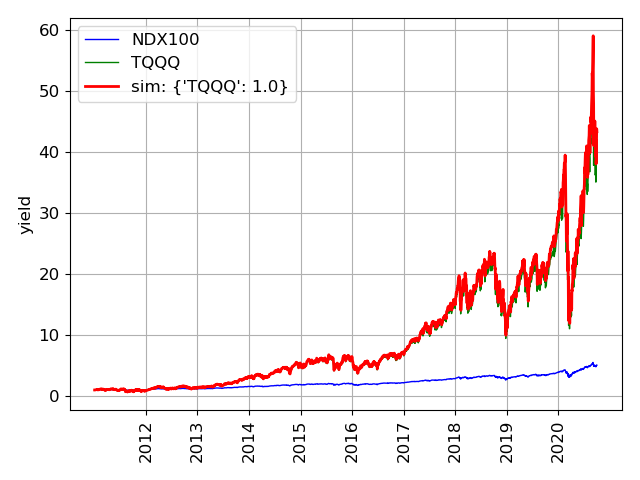}
\par\end{centering}
\caption{Data for the NDX100 index and TQQQ ETF for the 2011-2020 period. A
tax-free simulation of TQQQ (based on NDX100-TR) is plotted in red.}
\end{figure}

\subsection{Margin \label{subsec:Margin}}

The more traditional way to leverage (compared to LETFs) is borrowing
money to invest. This method is in principle better than of a daily
LETF, because it does not suffer as much from fluctuation of the underlying
index.

Interactive Brokers brokerage allows for a 2X leverage on margin (called
Reg T margin), meaning you can borrow against 100\% of your portfolio.
However, during the trading day itself the leverage can change dynamically.
If the portfolio value drops, the loan fraction of the total portfolio
increases which means higher leverage. IBKR allows for 4X leverage
(called maintenance margin) during the trading day, but at the end
of the day the original 2X is a maximum. Meaning, any deviation beyond
that limit will trigger a ``margin call'' which means the broker
will sell some of the portfolio to bring it back to the 2X leverage
limit \cite{ibkr_buy_power}. 

Our simulations only deal with end of day data, so we do not care
about the higher leverage allowed mid-day. Due to the fact that the
leverage changes dynamically, we will target 1.8X leverage, to leave
a $\sim10\%$ buffer to the maximum allowed leveraged. If the margin
leverage deviates by $10\%$ from the target, it will trigger a rebalance,
which is effectively ``margin calling'' yourself. The algorithm
for this is described in the next section.

The current cost of the loan in Interactive Brokers in 1.59\% (yearly).
So if the margin debt is $M$, each day the debt is increased by $M\cdot1.59/252$
percents.

We now compare the performace of the two methods of 2X leverage on
a backtest for the NDX100 index. For margin leverage, as the portfolio
value change the margin leverage $L$ will change dynamically, so
we define the target to be $L=2$ with 10\% allowed buffer prior to
rebalance (in practice 2X leaves no buffer, but we simulate this here
for fair comparison to the 2X LETF QLD). We also assume the margin
simulation is tax-free even though it is not possible to use this
method in a tax-free account (at least in Israel), again, for fair
comparion.

The two simulations side by side (the yield is defined as the total
portfolio size, minus the margin debt, which is roughly 100\% in the
case of 2X target leverage):

\begin{figure}[H]
\begin{centering}
\includegraphics[scale=0.7]{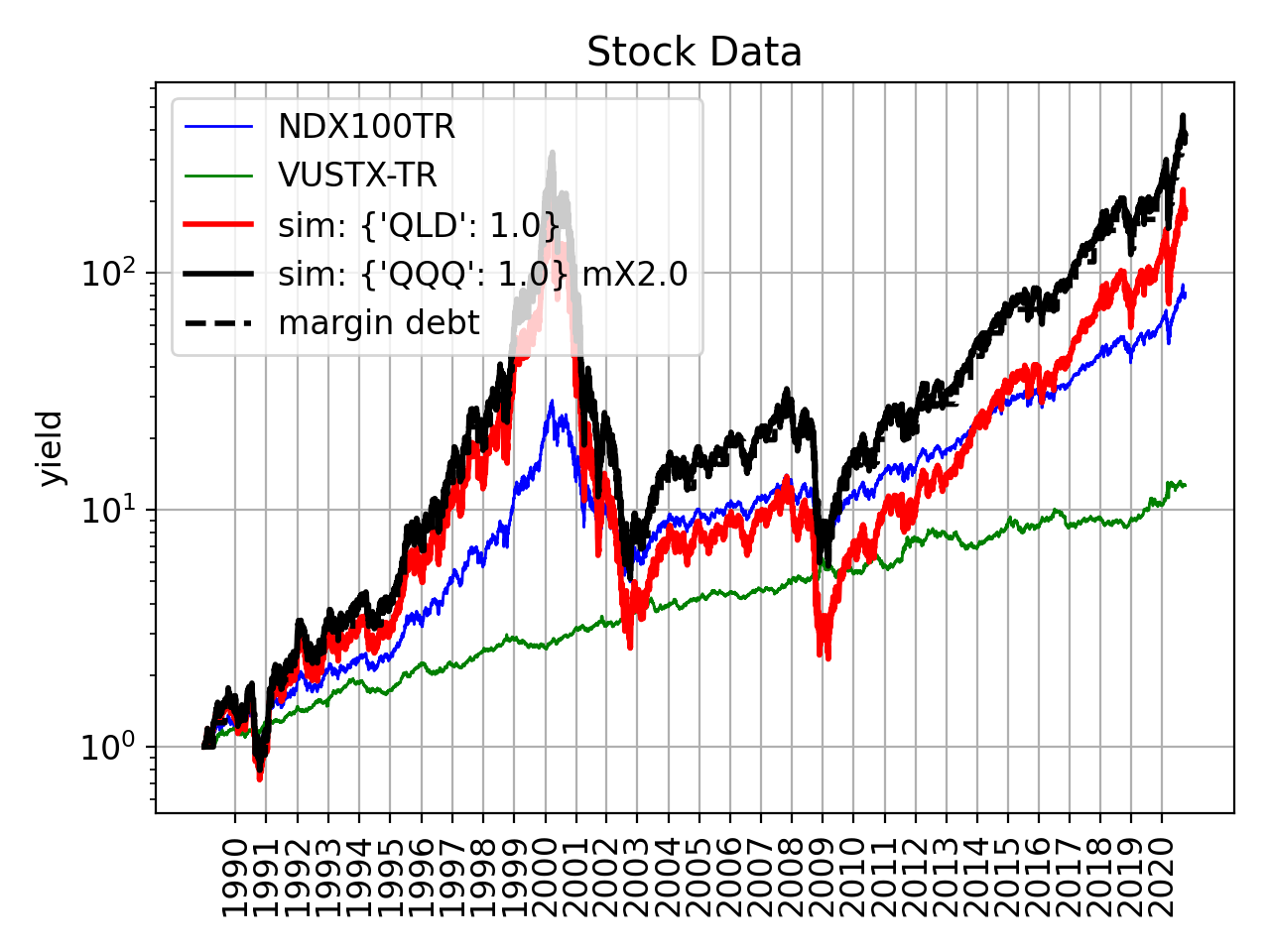}
\par\end{centering}
\caption{Tax-free simulations for 2X leverage on the NDX100 index for the period
1989-2020, using LETF (red) and margin (black). Margin debt is plotted
in dashed black.}
\end{figure}

The margin leverage as a function of time:

\begin{figure}[H]
\begin{centering}
\includegraphics[scale=0.5]{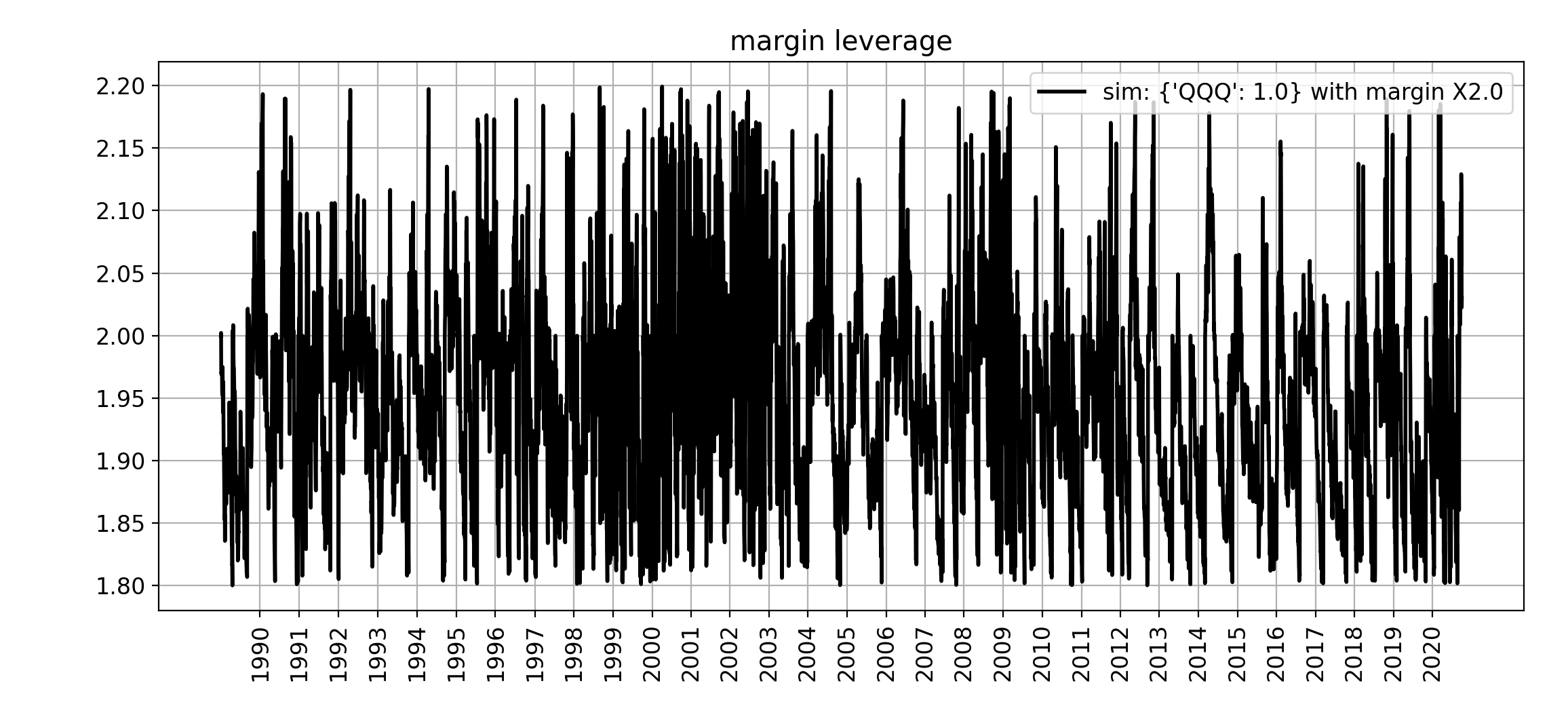}
\par\end{centering}
\caption{Evolution of the margin leverage in the 2X margin leveraged NDX100
simulation.}
\end{figure}
We can see the margin leverage $L$ changes dynamically but does not
exceed 10\% from the target due to active rebalancing, to be discussed
in section \ref{subsec:Rebalancing}.

Same for SP500 index:

\begin{figure}[H]
\begin{centering}
\includegraphics[scale=0.7]{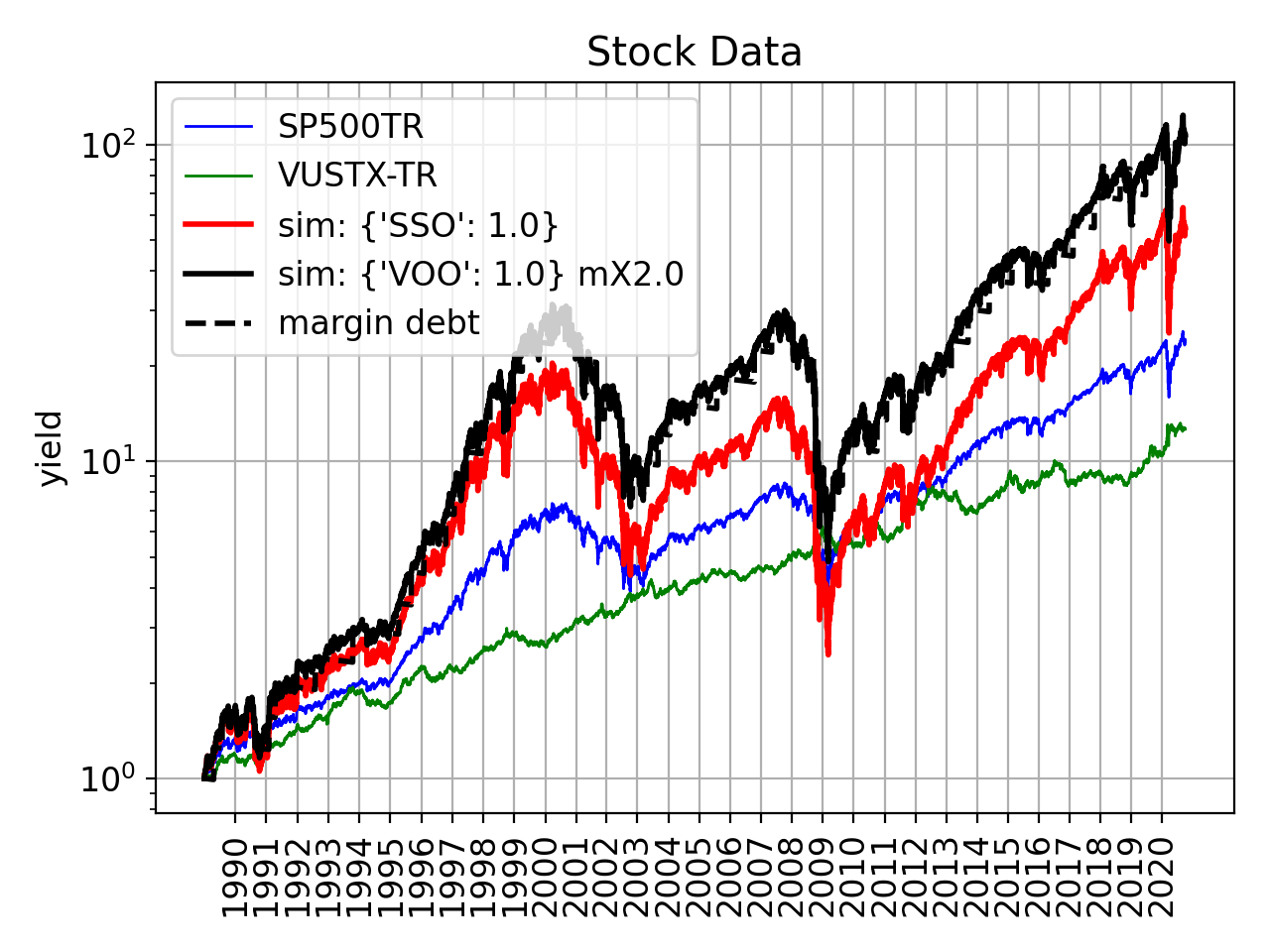}
\par\end{centering}
\caption{Tax-free simulations for 2X leverage on the SP500 index for the period
1989-2020, using LETF (red) and margin (black). Margin debt is plotted
in dashed black.}
\end{figure}

\subsection{Rebalancing\label{subsec:Rebalancing}}

So far we discussed the simulation of a single stock type. However,
a mixed portfolio of stocks and bonds can be beneficial, since the
two assets rise over the long term, yet somewhat anti-correlated,
so can cover over each other during large price swings. We define
a portfolio with some ideal (or target) fractions $f_{i}^{ideal}$,
and as is evolves to $f_{i}$ that deviates too much from the target,
we trigger a rebalance that will restore order. In our simulations
we picked the deviation trigger as 20\% for any stock (e.g. if we
pick $f=10\%$, it will trigger only at 30\%, and if $f=50\%$ it
will trigger at 30\% or 70\%).

Define the total portfolio values as $T$, and each fraction value
$T_{i}$ so therefore $f_{i}=T_{i}/T$. We want to transfer money
around $T_{i}'=T_{i}+\Delta T_{i}$ to rebalance the portfolio to
the ideal fractions:

\begin{equation}
f_{i}'=\frac{T_{i}'}{T}=f_{i}^{ideal}
\end{equation}

Therefore:

\begin{equation}
\Delta T_{i}=T\left(f_{i}^{ideal}-f_{i}\right)
\end{equation}

If $\Delta T_{i}>0$ we need to buy the stock, and if $\Delta T_{i}<0$
we sell. For a portfolio with two stock types (stocks and bonds) naturally
$\Delta T_{stocks}=-\Delta T_{bonds}$.

In the margin leverage scenario, we have margin debt $M$, so the
margin leverage is $L=\frac{T}{T-M}$. As we mentioned in section
\ref{subsec:Margin}, the margin leverage can change dynamically as
well, and if it deviates too much (we picked 10\% from $L^{ideal}$)
we also trigger a rebalance. If the ideal (or target) leverage is
$L^{ideal}$, then the loan might need to change to $M'=M+\Delta M$,
which also changes the total portfolio value $T'=T+\Delta M$:

\begin{equation}
L'=\frac{T'}{T'-M'}=\frac{T+\Delta M}{T-M}=L^{ideal}
\end{equation}

Therefore:

\begin{equation}
\Delta M=L^{ideal}\left(T-M\right)-T=T\left(\frac{L^{ideal}}{L}-1\right)
\end{equation}

If no margin debt is allowed then $M=0$ and of course $\Delta M=0$
always. 

During the rebalancing we can change both the component fractions
to the ideal fractions, and fix the margin leverage simultaneously:

\begin{equation}
f_{i}'=\frac{T_{i}'}{T'}=\frac{T_{i}+\Delta T_{i}}{T+\Delta M}=f_{i}^{ideal}
\end{equation}

Therefore:

\begin{equation}
\Delta T_{i}=T\left(f_{i}^{ideal}-f_{i}\right)+f_{i}^{ideal}\Delta M
\end{equation}

\subsubsection{Examples}

Backtest simulation of a 50\%/50\% non-leveraged VOO/VUSTX portfolio:

\begin{figure}[H]
\begin{centering}
\includegraphics[scale=0.7]{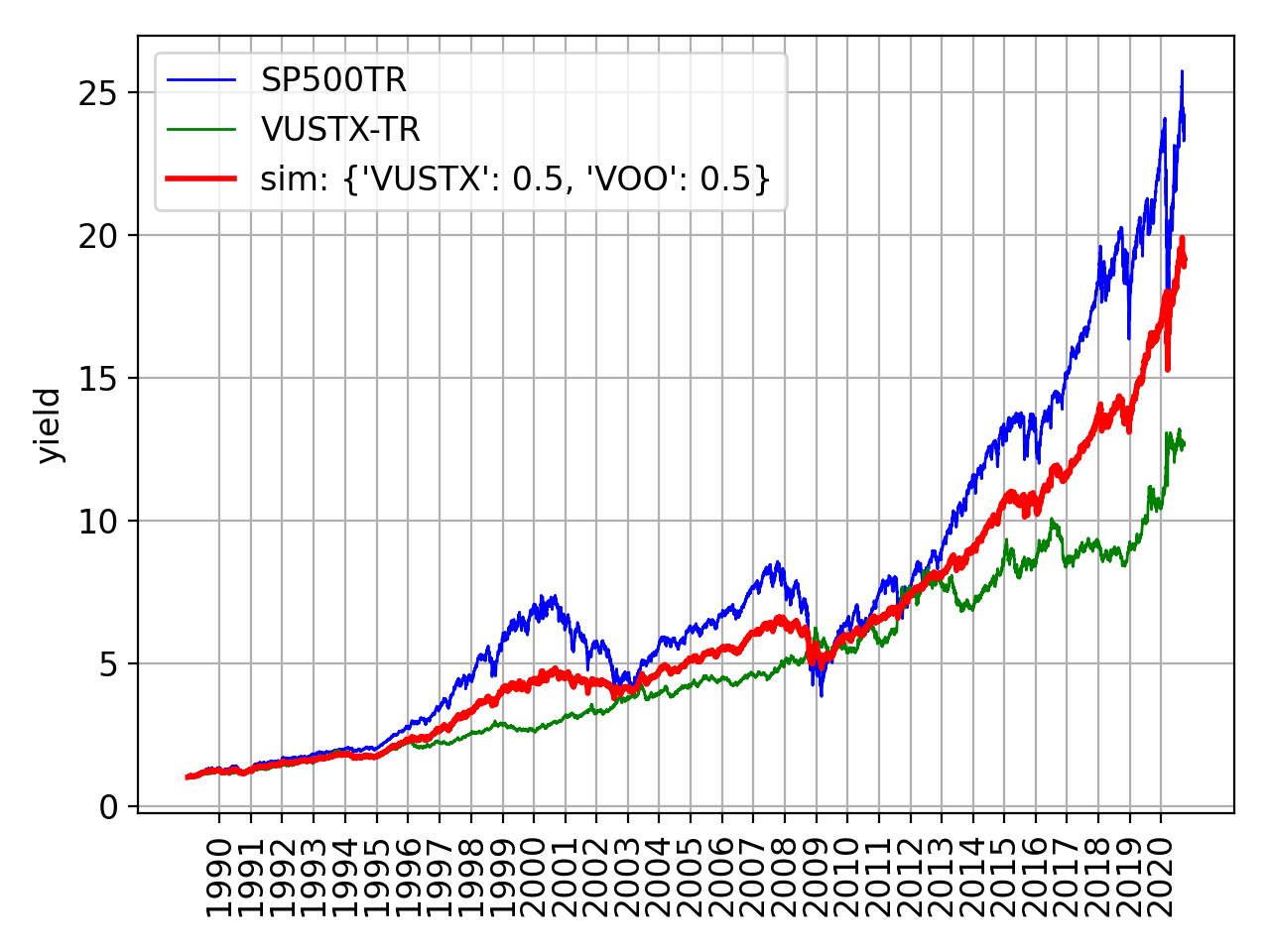}
\par\end{centering}
\caption{Tax-free simulation for a 50\%/50\% VOO/VUSTX portfolio in the period
1989-2020.}
\end{figure}

We can see intuitively that the stock component of the portfolio pulls
it upwards, and the bond component reduces drawdown.

The evolution of the fractions:

\begin{figure}[H]
\begin{centering}
\includegraphics[scale=0.7]{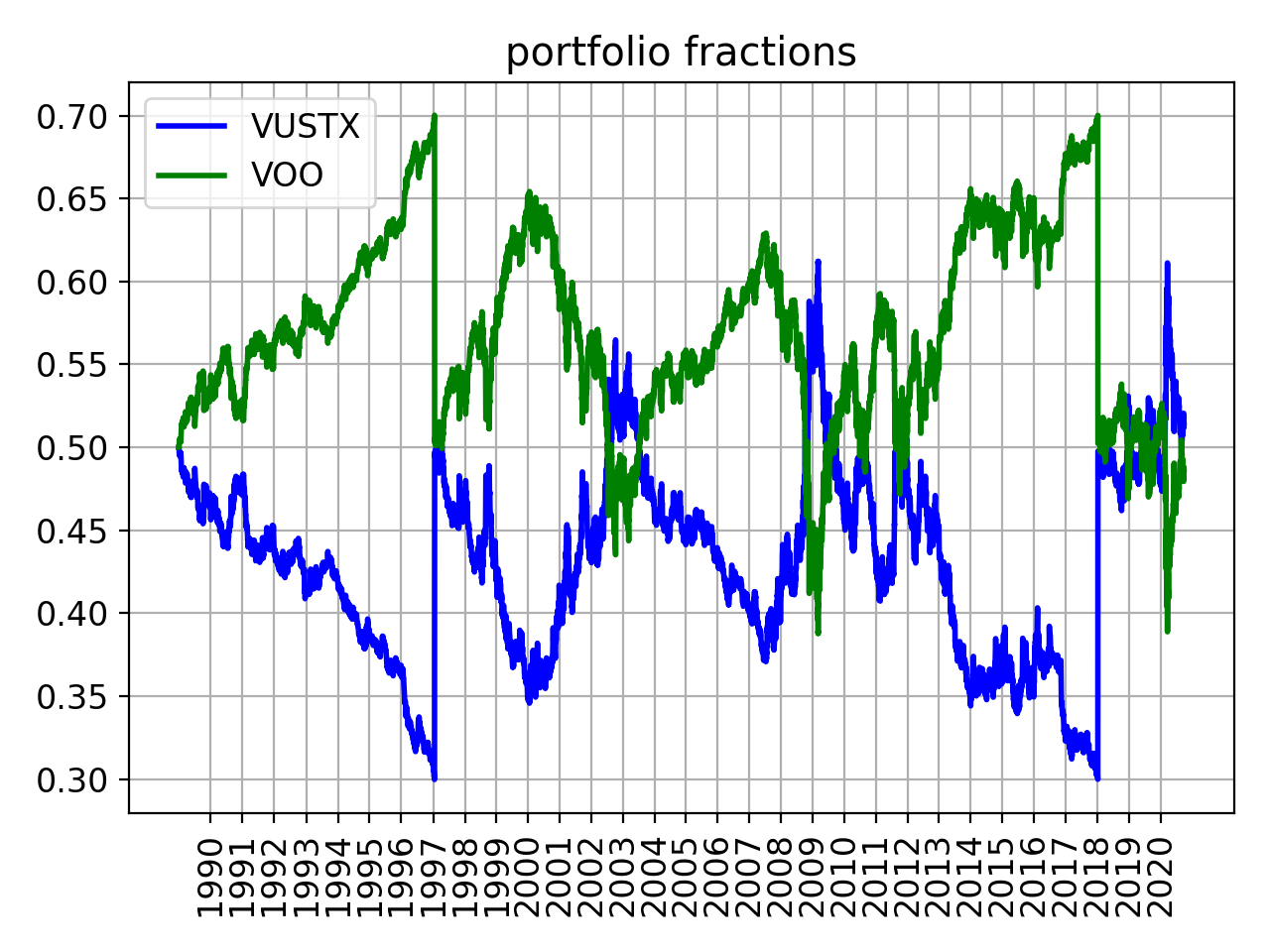}
\par\end{centering}
\caption{Evolution of the portfolio fractions of VOO and VUSTX in the 50\%/50\%
VOO/VUSTX simulation.}
\end{figure}

We can see it indeed changes and triggers rebalance at the correct
times.

Now move on to 2X levearged mixed portfolios, with both leverage methods.
For SP500 index:

\begin{figure}[H]
\begin{centering}
\includegraphics[scale=0.7]{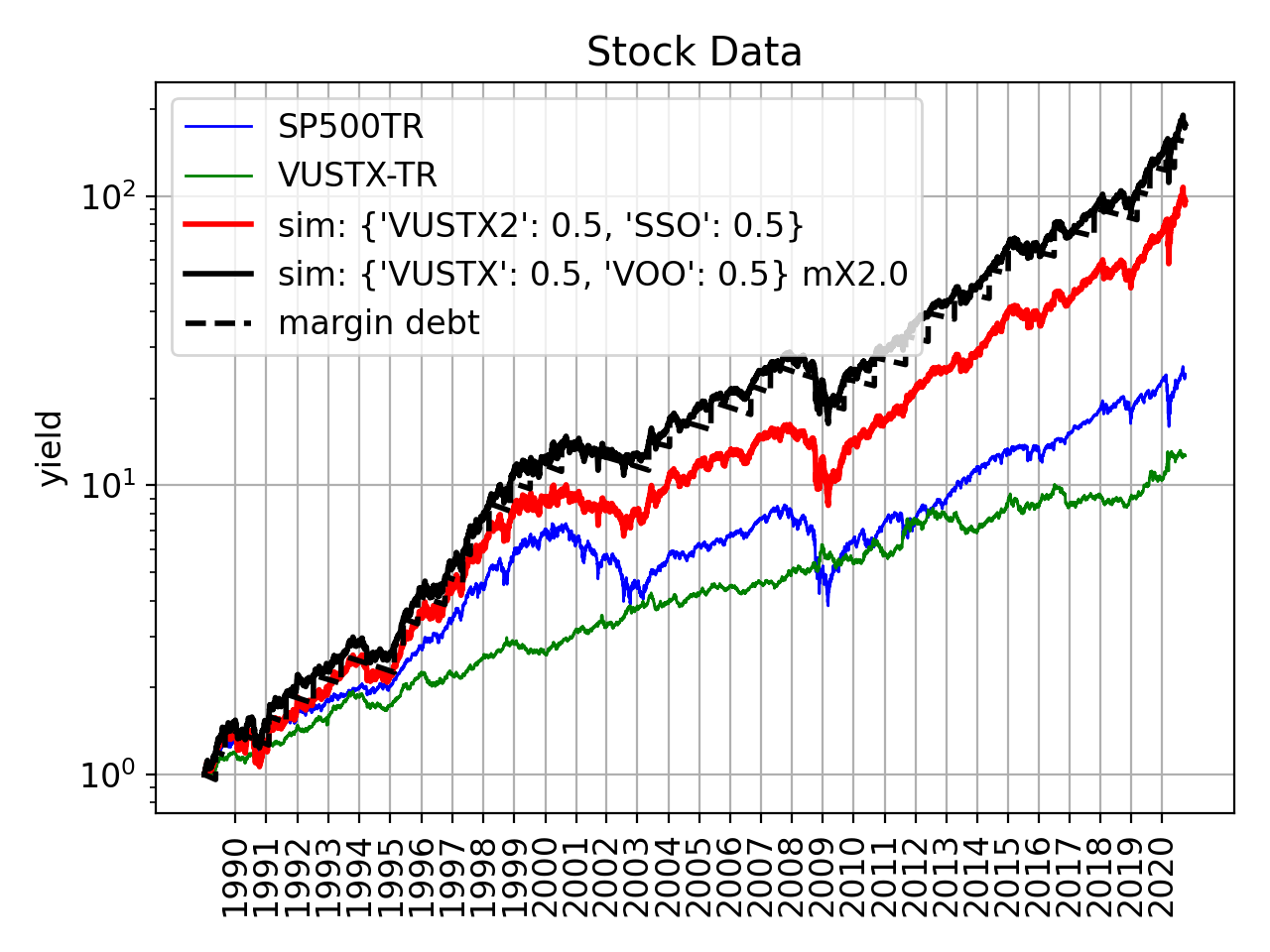}
\par\end{centering}
\caption{Tax-free simulation for a 50\%/50\% 2X leveraged SP500/VUSTX portfolios
in the period 1989-2020, using LETFs (red) and margin (black).}
\end{figure}

For NDX100 index:

\begin{figure}[H]
\begin{centering}
\includegraphics[scale=0.7]{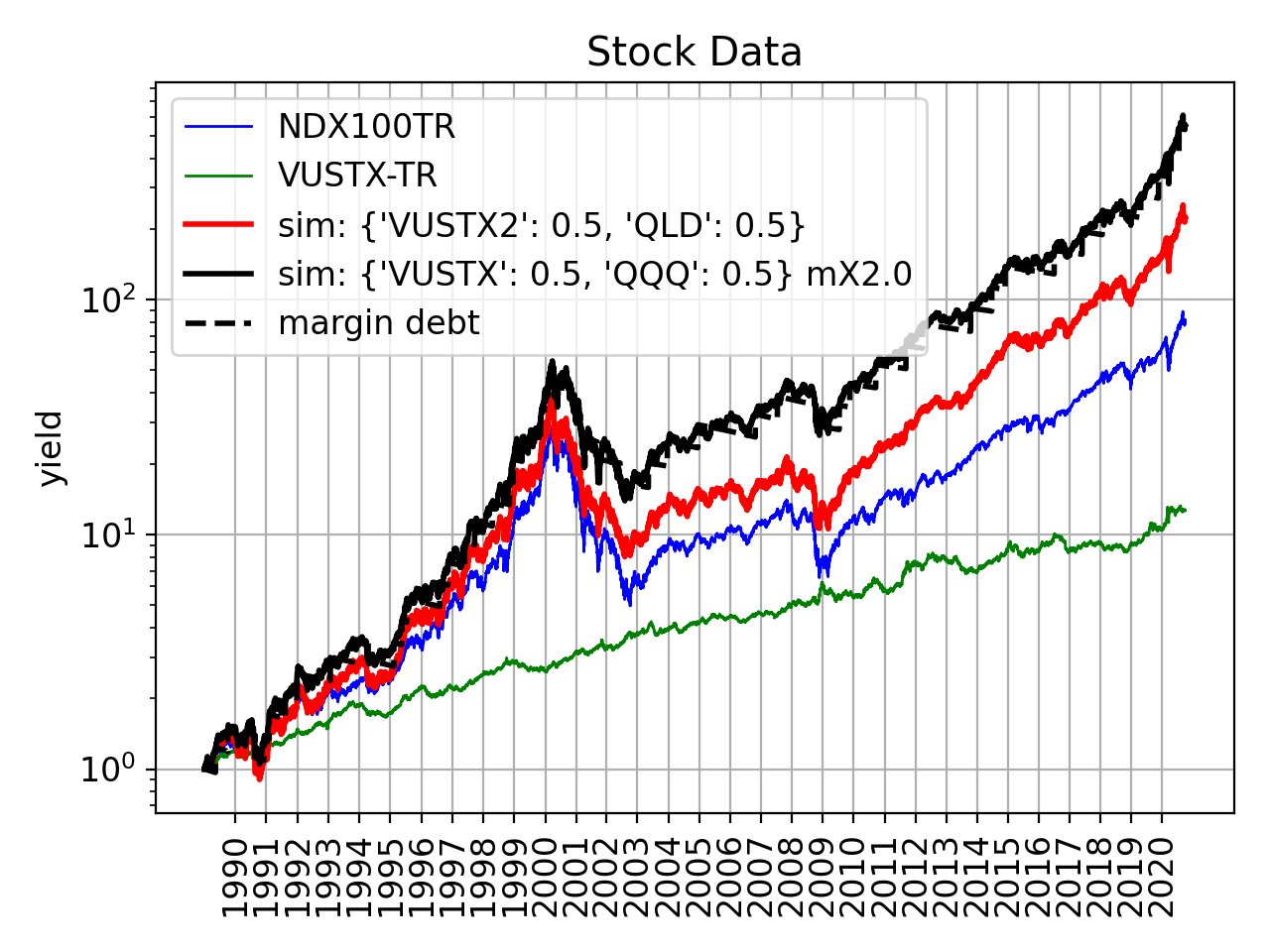}
\par\end{centering}
\caption{Tax-free simulation for a 50\%/50\% 2X leveraged NDX100/VUSTX portfolios
in the period 1989-2020, using LETFs (red) and margin (black).}
\end{figure}

We can see in this backtest the leveraged portfolio gave higher yields
than the non-leveraged portfolio, and also that the margin leverage
(at the given margin rate) is more effective than the LETFs, because
it is not vulnerable to daily fluctuations as much as LETFs.

\subsection{Capital Gains Tax}

In the previous section we saw margin leverage can be better than
LETF leverage. However, the simulations made assumed no taxes exist.
In reality (specific to Israel, at least) an investor with a tax-free
IRA account can use LETFs but cannot use margin leverage, so it is
only a theoretical exercise. In a normal taxable account, both leverage
strategies can be used. But for a taxable account, gains must be tracked
because capital gains tax needs to be paid for them. 

The tax needs to be paid only in the end of the year a paper is sold,
so the tax can be deffered for many years which is very beneficial.
In case the portfolio is composed of a single stock type, the investment
is simply buy-and-hold and there is no need to worry about taxes during
the simulation, except at the very end when we sell the entire portfolio.
But for a rebalanced portfolio with several stock types, we have to
sell papers from time to time, which is a negative effect that we
aim to quantify.

\subsubsection{Contributions to gains}

The gains $G$ need to be counted, and paid for in the end of the
year. Then we reset $G=0$ for the next year. However, if papers are
sold at a loss at the end of the year, the losses $G<0$ do not reset
but can be carried over to the following years to cancel out future
gains. 

Gains come from two sources: dividends, and paper sells. All dividends
count as gains so have to be tracked on a daily basis $\Delta G_{div}=D$.

Rebalancing requires selling papers. Consider a paper with value $V$
that has profit $P$ (can be negative in which case it is a loss),
the contribution to the gains depends on how much we need to sell
$\Delta T$. If $\Delta T>V$, then the paper is sold out completely,
the yearly gains increase by $\Delta G=P$, and we move on to the
next paper to complete the necessary sell amount. If $\Delta T\leq V$
then only part of the paper is sold, but the amount sold will count
as yearly gains. If the amount sold is larger than the profits then
the gains will max out by the profits $\Delta G=\text{sign}\left(P\right)\min\text{\ensuremath{\left(\Delta T,P\right)}}$.

To sum up:

\begin{equation}
\Delta G_{rebal}=\begin{cases}
\text{sign}\left(P\right)\min\text{\ensuremath{\left(\Delta T,P\right)}} & \Delta T\leq V\\
P & \Delta T>V
\end{cases}
\end{equation}

\subsubsection{Treating end of year taxes}

At the end of the year part of the portfolio needs to be sold to generate
the cash needed for the tax. If the gains are negative (loss) $G\leq0$,
nothing needs to be done, and the loss carries over to the following
year. 

For positive gains $G>0$, the capital gains tax fraction is $cgt=0.25$
(Israeli value), so the tax to pay is $t=G\cdot cgt$. At the end
of the year the portfolio is reduced to $T'=T-t$. 

However, as we sell some of the papers, they generate additional profit
or loss, which increase or decrease the total tax that is needed to
be paid. Define the profit for the portfolio to be $P$, so by definition
$P<T$. If we sell a portion $\Delta T$ the updated tax is therefore:

\begin{equation}
t'=\left(G+\Delta T\right)cgt
\end{equation}

$\Delta T$ itself is the total amount to be sold so $\Delta T=t'$.
The solution to the equation is:

\begin{equation}
\Delta T=t'=\frac{cgt}{1-cgt}G=\frac{G}{3}\equiv t^{*}
\end{equation}

So we can see the necessity to pay tax amplifies itself if we have
positive yearly gains $G>0$. However, as we said, this enlarged tax
is maxed out by the profitable part $P$ of the portfilio. If $t^{*}\leq P$
the solution above is valid, but after we sell off all the profit,
the remainder does not incur further tax. So for $t^{*}>P$ the tax
is paid for $G+P$:

\begin{equation}
\Delta T=t'=\left(G+P\right)cgt=\frac{G+P}{4}
\end{equation}

To sum up the different cases, as a function of gains $G$:

\begin{equation}
\Delta T=\begin{cases}
G\frac{cgt}{1-cgt} & G\leq\frac{1-cgt}{cgt}P\\
\left(G+P\right)cgt & G>\frac{1-cgt}{cgt}P
\end{cases}=\begin{cases}
\frac{G}{3} & G\leq3P\\
\frac{G}{4}+\frac{P}{4} & G>3P
\end{cases}
\end{equation}

For a given portfolio profit $P$, the amount to sell $\Delta T$
increases lineary with the yearly gains $G$, but once the sell amount
surpasses the profit $\Delta T=P$ it continues to increase linearly
with $G$ but with a reduced slope, which makes sense as we discussed. 

In the generalized case the portfolio is composed of many individual
papers and this needs to be done iteratively over all of them because
they will all be profitable or lossy to different extents. This brings
additional complications. 

In the equations above we assumed we have positive profit $P>0$,
but let us say we have total positive gains $G>0$ but the paper we
are currently selling is lossy $P<0$. We did not treat this case
before. In this case as we sell more it actually magnfies the loss,
the opposite of what we had before, which is beneficial. Meaning

\[
\Delta T=t'=\left(G-\Delta T\right)cgt
\]

The solution is

\begin{equation}
\Delta T=t'=\frac{cgt}{1+cgt}G=\frac{G}{5}\equiv t^{**}
\end{equation}

And again, the lossy sell is maxed out at $t^{**}=\left|P\right|$,
and for higher gains we have again:

\begin{equation}
\Delta T=t'=\left(G+P\right)cgt=\frac{G+P}{4}
\end{equation}

Again, summing up the solution but for $P<0$:

\begin{equation}
\Delta T=\begin{cases}
G\frac{cgt}{1+cgt} & G\leq\frac{1+cgt}{cgt}\left|P\right|\\
\left(G+P\right)cgt & G>\frac{1+cgt}{cgt}\left|P\right|
\end{cases}=\begin{cases}
\frac{G}{5} & G\leq5\left|P\right|\\
\frac{G}{4}+\frac{P}{4} & G>5\left|P\right|
\end{cases}
\end{equation}

The solution for both profitability cases:

\begin{align}
G^{*} & =\frac{1-\text{sign}\left(P\right)cgt}{cgt}\left|P\right|\\
\Delta T & =\begin{cases}
G\frac{cgt}{1-\text{sign}\left(P\right)cgt} & G\leq G^{*}\\
\left(G+P\right)cgt & G>G^{*}
\end{cases}\label{eq:delta_T_tax}
\end{align}

\begin{figure}[H]
\begin{centering}
\includegraphics[scale=0.7]{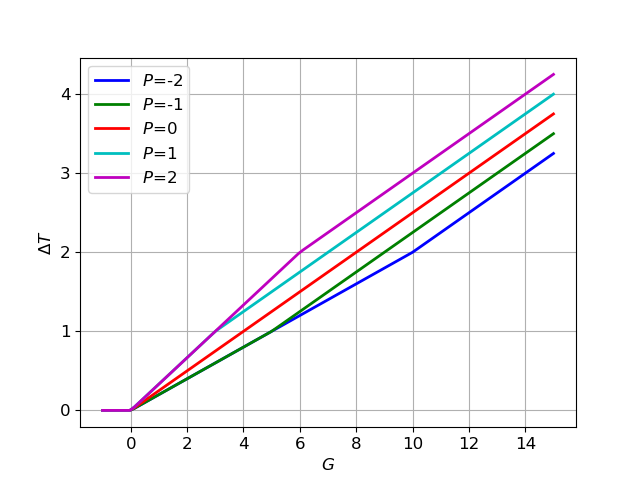}
\par\end{centering}
\caption{The amount $\Delta T$ that needs to be sold from a portfolio paper
with profit $P$, to pay for overall portfolio capital gains $G$,
as derived in Eq. (\ref{eq:delta_T_tax}).}
\end{figure}

If a specific paper we are looking at has value $V$ that is larger
than the total amount required to sell according to our equations
$V\geq\Delta T$, then all the tax necessary will be sold from that
paper alone and we are done. Otherwise, we sell all of it, and continue
to the next paper. 

When we sell a paper completely, the difference $\Delta T-V>0$ is
the tax that remains to be paid. So we continue to the next paper
with the gains updated to $G=\left(\Delta T-V\right)/cgt$.

The order we choose to sell the papers is what we call the tax scheme.
If we are free to choose which papers to sell, then the optimized
scheme will sort the papers by least profitable to most profitable,
and start selling the least profitable paper to minimize total paid
tax. If we are not free to choose the order of papers, such as in
Israeli brokers, we have to use the FIFO scheme, which sells the papers
by the order they were bought.

In the simulation engine we keep of track of all the papers of a specific
stock seperately. So when it is time to pay taxes for overall gains
$G$, we actually assign each stock type a separate gain $fG$ (remember
$\sum f=1$) and let each stock type to handle a separate chuck of
gains to pay taxes for. This is a simplifying choice that is made
due to our implementation of the simulation.

Summary of the algorithm:
\begin{enumerate}
\item We reach the end of year with gains $G$. If $G<0$ do nothing and
continue to the next year with the losses saved for the following
year.
\item If $G>0$, we need to pay tax. Each stock type is composed of a set
of papers, and is assigned gains $fG$ to handle.
\item Sort the papers by the chosen tax scheme, where each paper has value
and profit $\left\{ V_{i},P_{i}\right\} $.
\item Start at paper $i=1$.
\item Calculate the amount that needs to be sold to taxes $\Delta T$ using
Eq. (\ref{eq:delta_T_tax}).
\item If $\Delta T\leq V_{i}$ we sell an amount $\Delta T$ from paper
$i$ ($V_{i}'=V_{i}-\Delta T$), all taxes are fully paid. The procedure
is done and we reset $G=0$ for next year.
\item If $\Delta T>V_{i}$ we fully sell paper $i$ ($V_{i}'=0$), update
the gains to $G=\left(\Delta T-V\right)/cgt$
\item proceed to the next paper $i+1$ from step $\left(3\right)$.
\end{enumerate}
Addintional notes:
\begin{itemize}
\item If we are using margin leverage, we can pay the tax by simply borrowing
the required money and increasing the margin debt and margin leverage.
To avoid complicating the algorithm and have a uniform treatment for
both margin and margin-free strategies, we will sell papers to pay
the tax, in the same method described in this section. This means
the total portfolio size is reduced while the debt does not so the
margin leverage increases in the process (until the next rebalancing).
\item In any rebalanced taxable portfolio, but especially in a leveraged
one, it is possible to reach a situation where rebalances during the
year generated some gains, and then the market crashes so severly
that at the end of the year selling the entire portfolio might not
generate enough cash to pay off the tax. In that case you end up in
a problem with the tax authority, so you have to take external loans
or flee the country. 
\end{itemize}

\subsubsection{Examples}

\paragraph{Example 1}

Backtest simulation for 50\%/50\% VUSTX/VOO with taxes in the optimzed
scheme:

\begin{figure}[H]
\begin{centering}
\includegraphics[scale=0.6]{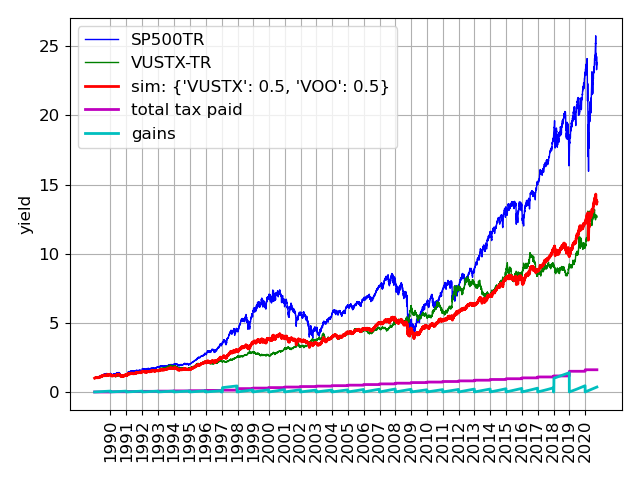}
\par\end{centering}
\caption{Taxed simulation for 50\%/50\% VOO/VUSTX for the period 1989-2020.
Except for the portfolio value (red) we also plot the tracked gains
(cyan) and total paid taxes (purple).}
\end{figure}

As usual, in the plot we have the data for SP500-TR and VUSTX-TR,
as well as the total portfolio value in the simulation. In addition,
we track the gains as a function of time, and the cummulative taxes
paid for those gains at the end of each year. The gains (or losses)
are generated upon selling of papers during rebalances, or due to
dividends. The ``sawtooth'' form of the gains mainly comes from
dividends that accumulate during the year, and reset at the of each
year.

\paragraph{Example 2}

Backtest simulation for 50\%/50\% VUSTX3/UPRO (3X LETFs of bonds and
SP500), with taxes in the optimzed scheme:

\begin{figure}[H]
\begin{centering}
\includegraphics[scale=0.6]{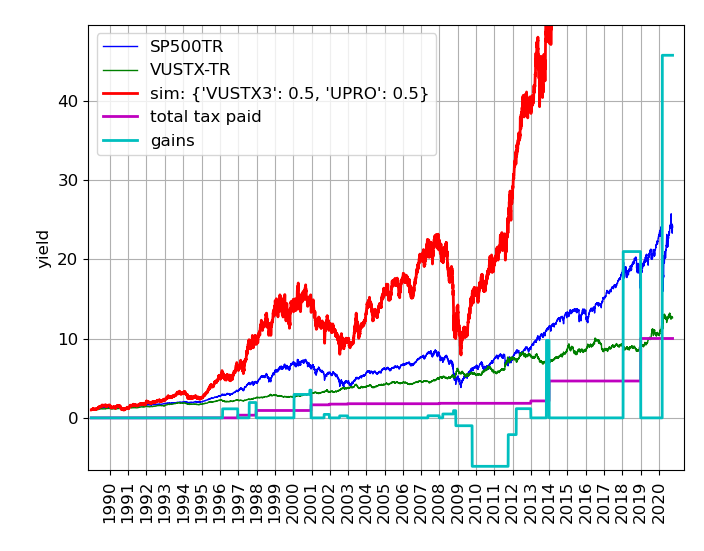}
\par\end{centering}
\caption{Taxed simulation for 50\%/50\% 3X leveraged SP500/VUSTX for the period
1989-2020. Except for the portfolio value (red) we also plot the tracked
gains (cyan) and total paid taxes (purple).}
\end{figure}

Since we consider LETFs where we assume the dividends are negligible,
the gains/losses are generated solely due to rebalances. In the backtest
above no rebalances were triggered until 1996, so there were no gains
registered. The plot is focused on the gains so we do not see the
final portfolio yield which is around $\sim200$ at the end of the
simulation period, equivalent to $\sim18\%$ compound annual growth
rate (CAGR). 

Note that when the gains are negative, they are transferred year to
year. In turns out that in this case using the FIFO tax does not give
a very different outcome.

\paragraph{Example 3}

Backtest of a 50\%/50\% VUSTX3/TQQQ (3X LETFs of bonds and NDX100):

\begin{figure}[H]
\begin{centering}
\includegraphics[scale=0.6]{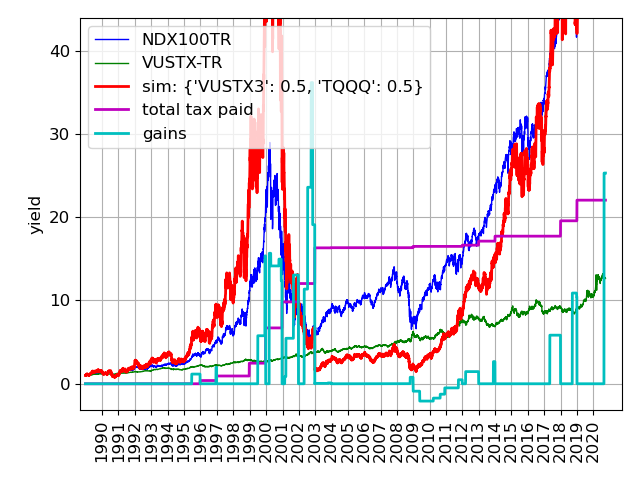}
\par\end{centering}
\caption{Taxed simulation for 50\%/50\% 3X leveraged NDX100/VUSTX for the period
1989-2020. Except for the portfolio value (red) we also plot the tracked
gains (cyan) and total paid taxes (purple).}
\end{figure}

We can see in this example that taxes can hit really hard. The market
surge at the year 2000 triggered many rebalances which in turn generate
gains. These massive gains require tax to be paid for them, even though
the market came crashing down after than, causing the portfolio to
shrink. 

It important to note that the simulation outcome is highly sensitive
to the rebalancing trigger. As we discussed in section \ref{subsec:Rebalancing},
we trigger a rebalance at 20\% deviation from the 50\%/50\% fractions.
Example for the evolution at different rebalancing deviation triggers:

\begin{figure}[H]
\begin{centering}
\includegraphics[scale=0.6]{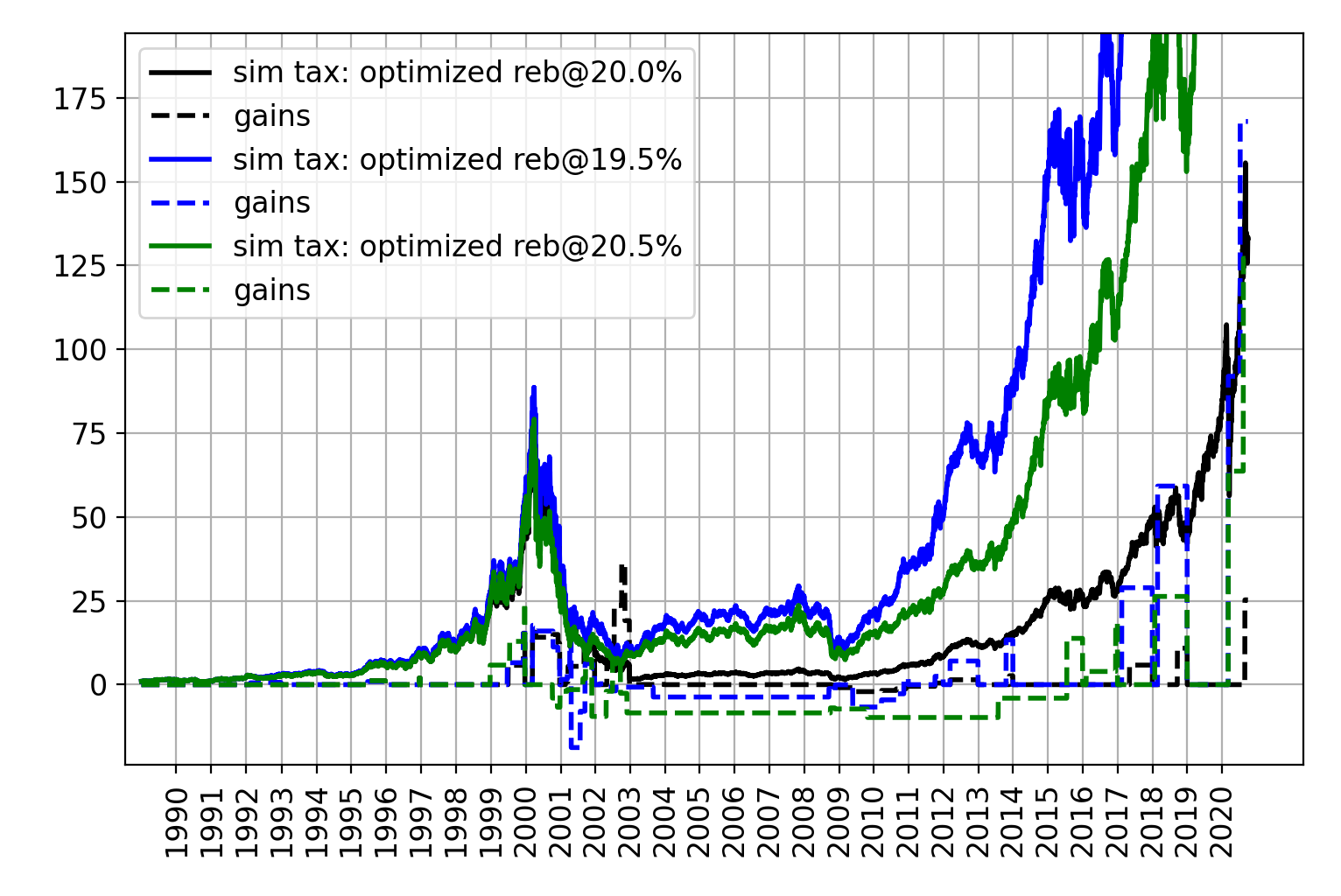}
\par\end{centering}
\caption{Taxed simulations for 50\%/50\% 3X leveraged NDX100/VUSTX for the
period 1989-2020, with slightly different rebalancing triggers.}
\end{figure}

If the simulation were tax-free, there would still be differences,
but not as extreme.

Backtests are nice, but we cannot depend on them when comparing different
strategies for the future. As was just demonstrated, even arbitrary
parameters such as the rebalance trigger can vastly impact the result.

\section{Bootstrapped Monte-Carlo Simulations \label{sec:Monte-Carlo-simulations}}

As discussed in the introduction, we predict the yield of a given
portfolio in a probabalistic way by assuming the daily price changes
of the past represent the future.

\subsection{Method}

In the bootstrapped Monte-Carlo (BMC) method, we generate possible
realizations of multiple year periods of stock histories by stitching
together daily price changes (randomly sampled with repetitions) from
the past data (defined in chapter \ref{sec:Data}). Consectutive days
in the stock market are mostly uncorrelated (the efficient market
drives it to this state), but when we sample days from the past we
do it in batches of 5 consecutive days to preserve any real correlations
that do appear in the real stock market (sampling single days changes
the results but not in a meaningful way). Also, we want to preserve
correlations between different assets in the same day (specifically
the very important anti-correlation of stocks and bonds) so when we
samples a random day from the past, we take the price changes for
both stocks and bonds from that day (and also the LIBOR rate, for
good measure).

Examples of several random 10-year realizations of the SP500:

\begin{figure}[H]
\begin{centering}
\includegraphics[scale=0.6]{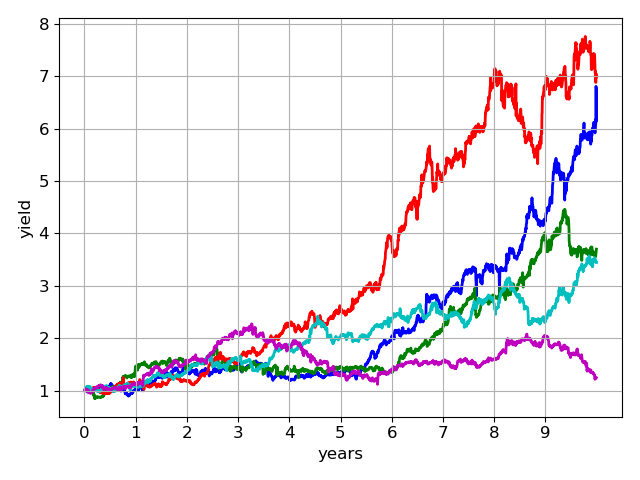}
\par\end{centering}
\caption{Synthetic realizations of 10-year evolutions of the SP500 index, based
on the 1989-2020 data.}
\end{figure}

\subsection{Risk/Reward Metrics \label{subsec:Risk/Reward-metrics}}

Define the porfolio yield as a function of time as $y\left(t\right)$.
A long term passive investor needs to define the time horizon of the
investment $t_{end}$. During that time the investment should be considered
locked, ``black box'', to be accessed only in the end. Therefore,
given the porfolio final yield probability distribution $y\left(t_{end}\right)$,
the statistical measure for reward metric will be defined as the 50\%
percentile of the distribution (median value), and the risk metric
defined as the 5\% percentile.

Mostly in the literature, risk is measured during the investment period
itself as the volatility of $y\left(t\right)$ (e.g. Sharpe ratio).
This is not a rational risk metric, but a psychological risk metric.
Nevertheless, we calculate two psychological risk metrics for comparison.
One will be the (5\% percentile of the) minimal yield during the investment
period $\min_{t}\left(y\left(t\right)\right)$ (how low will the portfolio
get if you constantly monitor it), and the second will be the (50\%
percentile of the) maximal drawdown (maximal drop of the portfolio
from any new high point reached during a portfolio's evolution) \cite{drawdown}.

As an example, we performed 2000 realizations of 10-year simulation
of several portfolios, and saved the outputs above for each of them.
The portfolios are 100\% SP500 with 1X, 2X and 3X leverage. A histogram
of the resulting yields $y\left(t_{end}\right)$ (left subplot), minimal
yields $\min_{t}\left(y\left(t\right)\right)$ (middle subplot) and
maximal drawdowns (right subplot).
\begin{center}
\begin{figure}[H]
\begin{centering}
\includegraphics[scale=0.43]{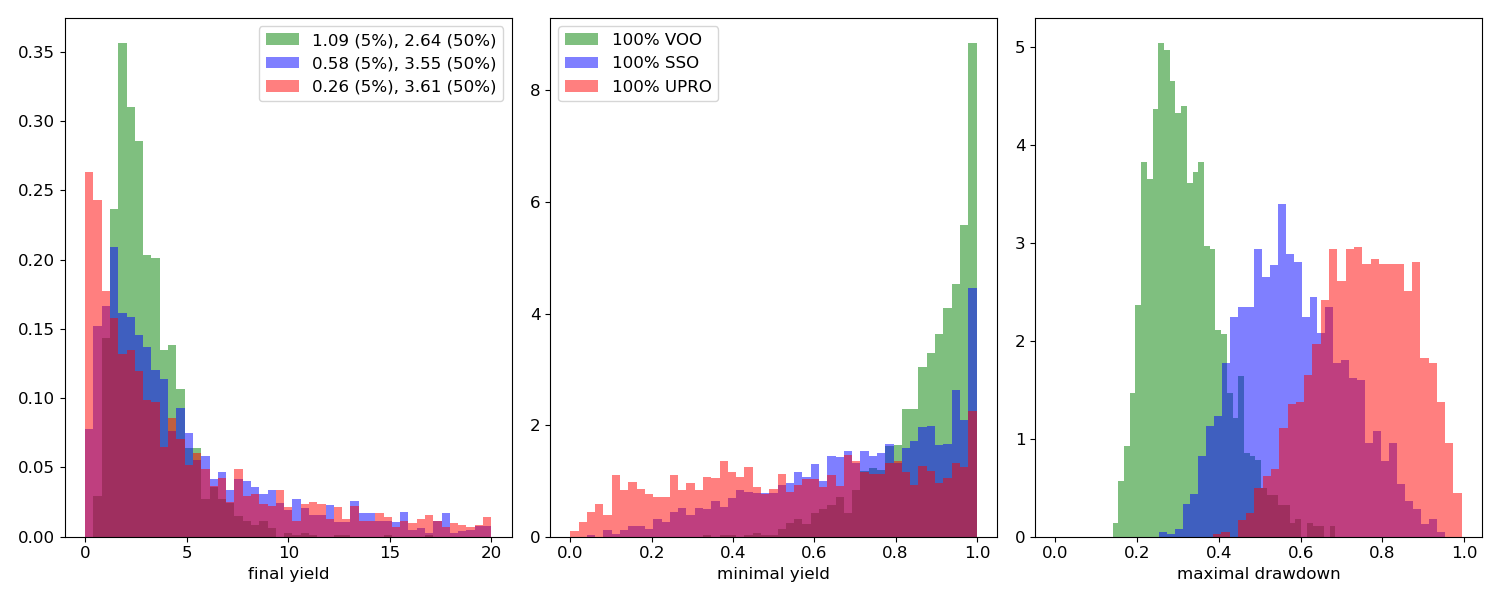}
\par\end{centering}
\caption{Probability distributions of the final yield, minimal yield and maximal
drawdown, generated using BMC for portfolios of 100\% SP500 with 1X,
2X and 3X leverage.}
\end{figure}
\par\end{center}

The legend in the middle subplot indicates the color for each portfolio.
On the left plot we have the yield distribution for each portfolio.
We can see that increasing the leverage elongates the high yield tail
of the distribution, but the low yields are getting heavier as well.
The percentiles of the yield distributions are what matters, and the
legend indicates the risk/reward metrics (5\% and 50\% percentiles).
We see that increasing the leverage increases both the (rational)
risk and the reward. Investing for a 10 year period in the unleveraged
portfolio gives less than 5\% chance for ending up with less money
than was invested (we do not take inflation into account). With the
leveraged portfolio, the chance for ending up with less money is much
higher. Looing at the two other plots of the psychological risk metric,
we see that they too give higher risk to the leveraged portfolios.

Next, we repeat the analysis for 50\%/50\% portfolios of SP500 and
bonds, with 1X, 2X and 3X leverage:
\begin{center}
\begin{figure}[H]
\begin{centering}
\includegraphics[scale=0.43]{pics/hists_MC_sp500_vustx_10years_tax_free}
\par\end{centering}
\caption{Probability distributions of the final yield, minimal yield and maximal
drawdown, generated using BMC for portfolios of 50\%/50\% portfolios
of SP500 and bonds, with 1X, 2X and 3X leverage.}
\end{figure}
\par\end{center}

Now the picture is much different. The mixed unleveraged portfolio
has about the same reward as before, but with reduced risk. Increasing
leverage boosts the reward, and surprisingly the risk almost does
not change (increases but much less than before). The mixed 3X leveraged
portfolio beats the nonleveraged 100\% SP500 portfolio on both risk
and reward, by a vast amount. The leverage does increase the psychological
risk metrics, meaning the investment is a much more bumpy ride, but
the (rational) end result is much better.

\subsection{Numeric Uncertainty}

Since we do not have the analytic forms of the probability distribution
of the final yield (or the other metrics from section \ref{subsec:Risk/Reward-metrics}),
but only finite size samples of them, the calculated risk/reward metric
have a numeric uncertainty to them. We quantify this uncertainty using
bootstrap: the Monte-Carlo sampling yielded a total of $N=2000$ samples
$\left\{ y\right\} _{i=1}^{N}$ that we have to work with. We perform
another Monte-Carlo where in each realization we pick $N$ random
samples (with possible repetition) out of $\left\{ y\right\} _{i=1}^{N}$
and recalculate the metric we are interested in $y_{metric}$ (such
as the reward metric which is 50\% percentile). Repeating this $M=300$
times we end up with a distribution of those metrics $\left\{ y_{metric}\right\} _{i=1}^{M}$
and we define the uncertainty in $y_{metric}$ as the 32\%-68\% confidence
interval (analogoues to a $\pm1\sigma$ interval). This interval is
what what we draw for each portfolio in section \ref{subsec:Results}.
As $N$ increases the numeric uncertainty decreases, and for our purposes
the calculated uncertainty is low enough to reach meaningful conclusions.

\subsection{Results \label{subsec:Results}}

In this section we repeat the Monte-Carlo calculations from section
\ref{subsec:Risk/Reward-metrics} for a variety of portfolios with
different fractions of stocks and bonds (5\% intervals between 0\%
and 100\% stocks). Each color in the following plots deals with two
different stock types, where the number on the top right of each point
indicates the percent of stocks within the portfolio. The portfolio
combinations we explore are SP500/bonds and NDX100/bonds, for an unleveraged,
2X and 3X LETFs portfolios, and 1.8X margin leveraged portfolio. In
the tax-free simulations the margin leverage is purely theoretical
since it cannot be implemented (at least in Israeli IRA accounts).

All plots share the same y-axis which is the reward metric, on the
left y-axis it is the final yield 50\% percentile after the investment
period, and on the right y-axis it is the associated compound annual
growth rate (CAGR). The x-axis of each plot is a different risk metric,
on the left plot it is the final yield 5\% percentile (the rational
risk metric), in the middle plot it is the minimal yield 5\% percentile,
and on the right plot it is the maximal drawdown 50\% percentile.

Our benchmark investment period will be 10 years, and we compare the
outcomes between tax-free and taxed investments. We then follow up
with the outcomes for 5 and 20 years tax-free investments, as well
as a ``bearish'' scenario.

\subsubsection{Tax free (10 years)}
\begin{center}
\begin{figure}[H]
\begin{centering}
\includegraphics[scale=0.45]{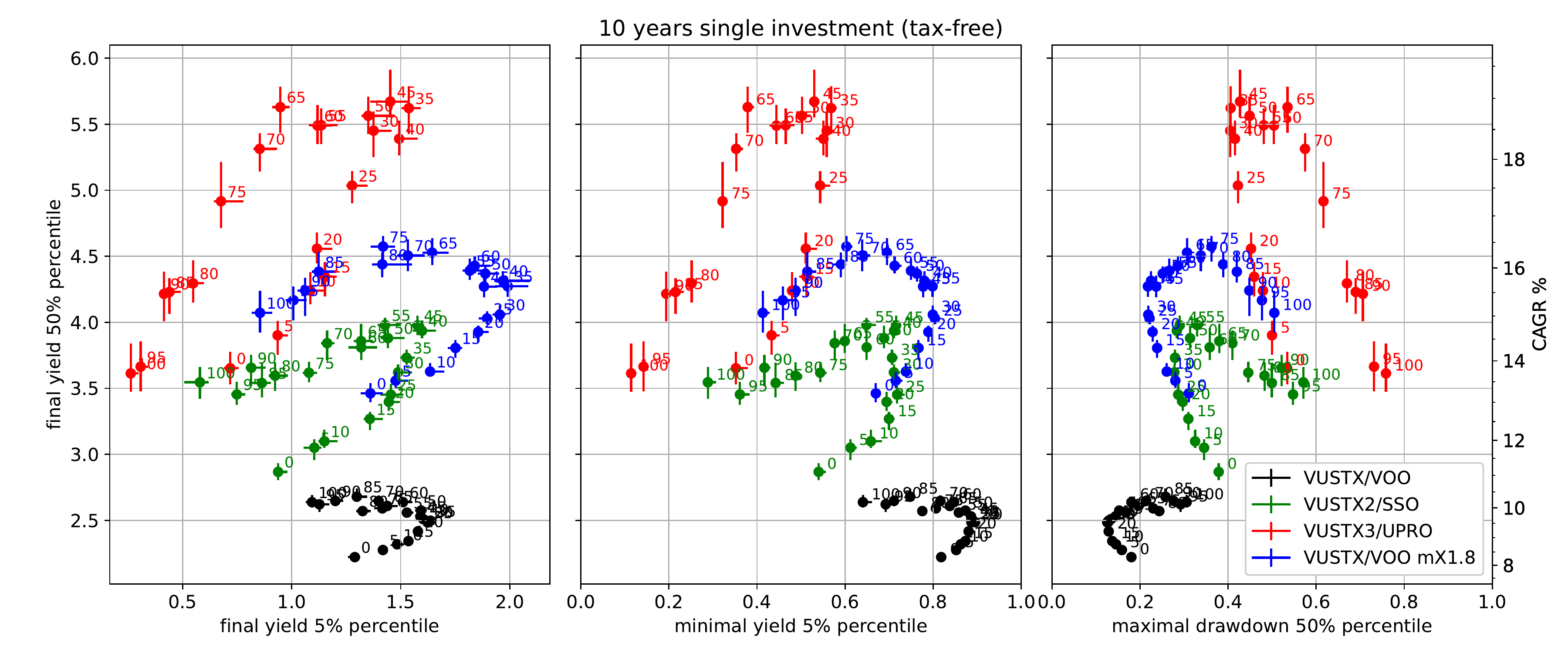}
\par\end{centering}
\caption{SP500/bonds portfolio metrics for a 10 years tax-free investment.}
\end{figure}
\par\end{center}

\begin{center}
\begin{figure}[H]
\begin{centering}
\includegraphics[scale=0.45]{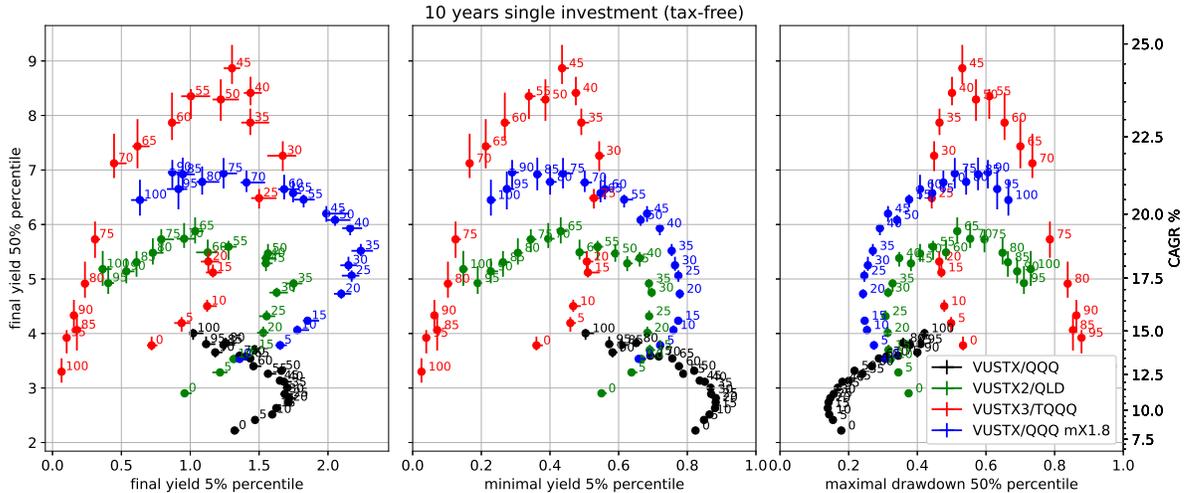}
\par\end{centering}
\caption{NDX100/bonds portfolio metrics for a 10 years tax-free investment.}
\end{figure}
\par\end{center}

First we examine the unleveraged portfolios (in black). We can see
that a 100\% stock portfolio has a higher risk and reward compared
to a 100\% bond portfolio. The mixed portfolio does not lie on the
interpolation line between these two but rather has a higher risk-adjused
reward \cite{WikiMarkowitz}. The NDX100 gives higher reward than
the SP500, for roughly the same risk. The least risky portfolio (measured
by the rational risk metric) is somewhere in the range of 25\%-45\%
stocks allocation.

Moving on to the leveraged portfolios, we can see dramatically higher
rewards in the mixed portfolios. Comparing the least risky portfolios
in each set, we see that LETF portfolios (in green and red) do increase
the (rational) risk but not in a significant way. If our benchmark
were a 100\% unleveraged stock portfolio, the leveraged mixed portfolios
can outperform it at the same risk. Examining the psychological risk
metrics, we can see the leveraged portfolios increase those metrics
more significantly.

It is also interesting to compare the 100\% stock portfolios in terms
of reward. For the SP500 the 3X leveraged portfolio beats the unleveraged
portfolio, but it is the opposite for NDX100. This is (probably) because
the NDX100 data has higher volatility. Both cases are horrible in
terms of risk. It is quite remarkable how adding bonds to a leveraged
portfolio turns the conclusions on their head.

The margin leveraged portfolio (in blue) is interesting because it
allows to outperform the unleveraged portfolio in both risk and reward.
This is because of its abillity to overcome small daily fluctuations,
unlike the daily-leveraged ETFs. We note again that for a tax-free
account (in Israel) this is not a possible portfolio anyway.

For a long term (10 year) investor happy with the (rational) risk
of 100\% stocks (SP500 or NDX100) unleveraged portfolio, switching
to a 3X leveraged portfolio with around 35\%-45\% stocks allocation
can significantly boost the reward, while reducing the (rational)
risk at the same time. If the same investor does not wish to take
on more psychologal risk than what the 100\% unleveraged stocks portfolio
offers, then switching to a 2X leveraged portfolio (40\%-60\% SP500
allocation and 35\%-45\% NDX100 allocation) can deliver higher reward
with lower risk (both rational and psychological).

\subsubsection{With taxes (10 years)}
\begin{center}
\begin{figure}[H]
\begin{centering}
\includegraphics[scale=0.45]{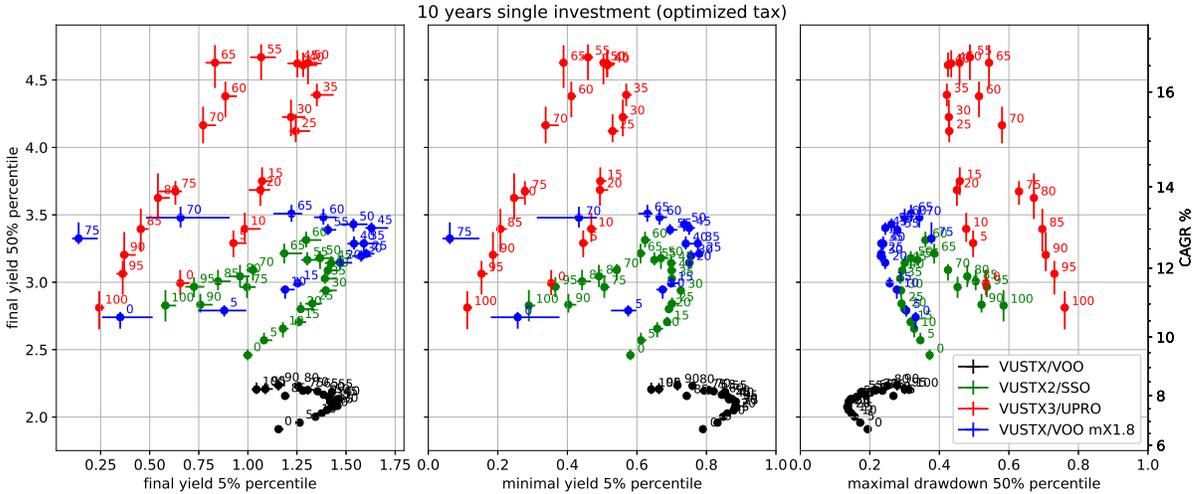}
\par\end{centering}
\caption{SP500/bonds portfolio metrics for a 10 years taxed investment.}
\end{figure}
\par\end{center}

\begin{center}
\begin{figure}[H]
\begin{centering}
\includegraphics[scale=0.45]{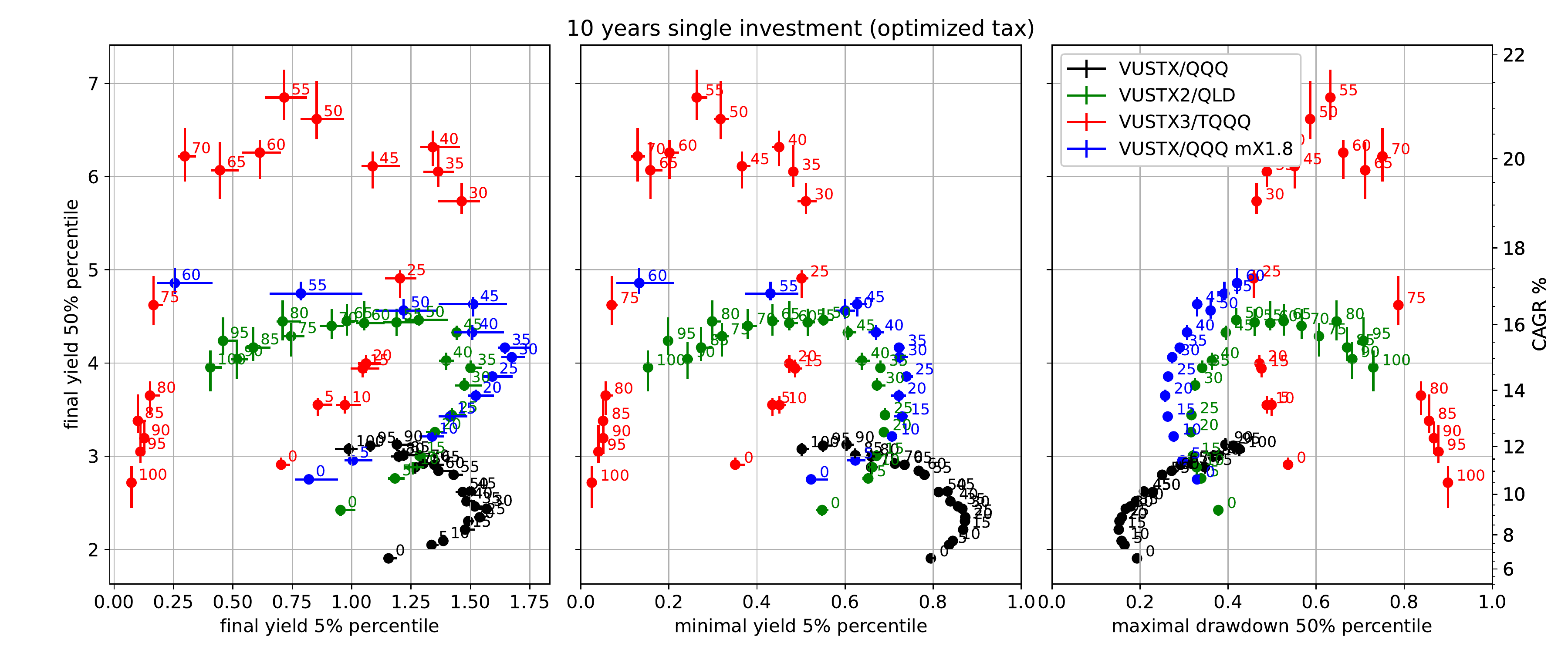}
\par\end{centering}
\caption{NDX100/bonds portfolio metrics for a 10 years taxed investment.}
\end{figure}
\par\end{center}

We can see that taxes reduce the reward (and increase the risk), but
if one wishes to invest in a taxable account, it is still meaningful
to ask how leverage can improve the performance of the portfolio.
We can see that the benefits of leverage using LETFs did not change,
meaning that the tax events generated by rebalancing were not significant
(compared to the unleveraged case). On the other hand, taxes significantly
hurt the performace of margin leverage. For high stock allocation
(over 80\% SP500 or over 60\% NDX100) the risk is so high than the
yield can even go negative (capital gains taxes need to be paid but
the portfolio is null). In lower stock allocations 10\%-50\% the 1.8X
margin leveraged portfolios beats the 2X LETF portfolio on both reward
and risk (rational and psychological).

The conclusions from the previous section on the benefits of the 3X
leveraged portfolio remain unchanged, even when taxes are taken into
account.

\subsubsection{Tax free (5 years)}
\begin{center}
\begin{figure}[H]
\begin{centering}
\includegraphics[scale=0.45]{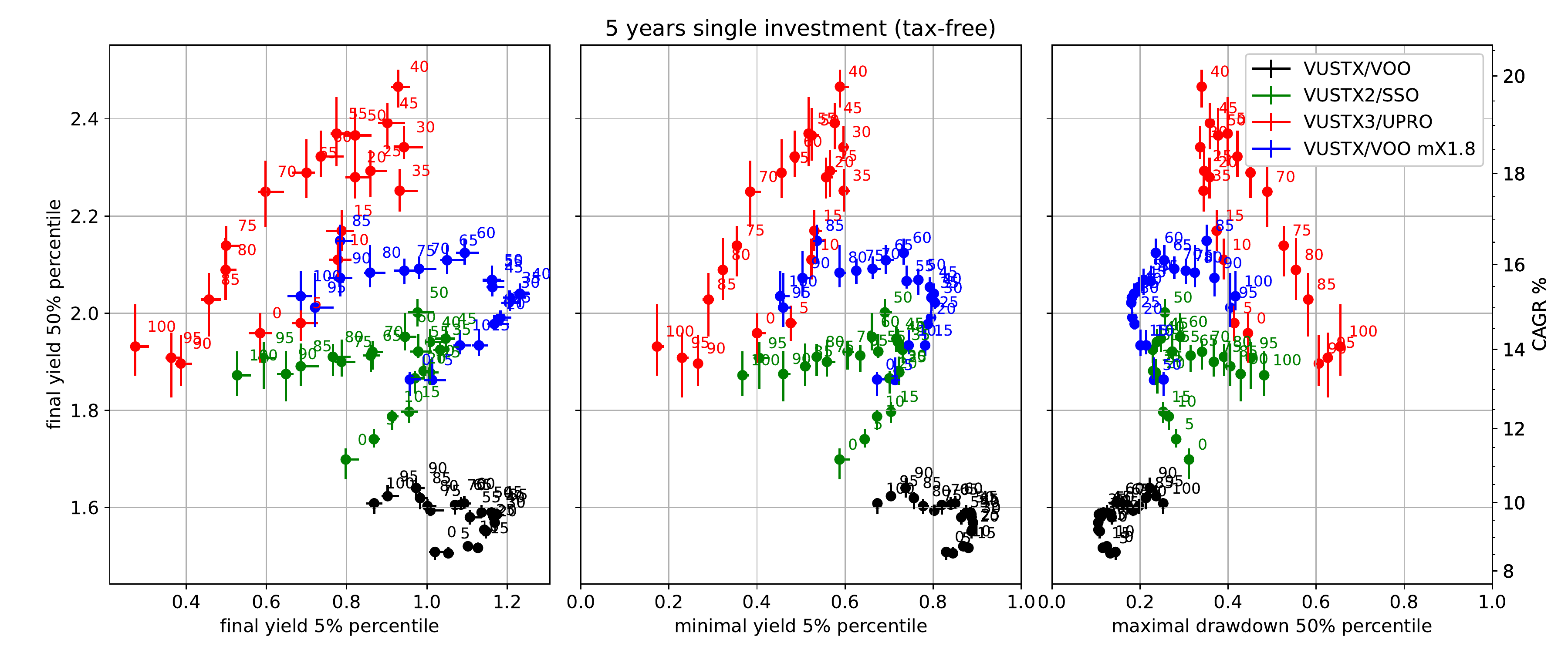}
\par\end{centering}
\caption{SP500/bonds portfolio metrics for a 5 years tax-free investment.}
\end{figure}
\begin{figure}[H]
\begin{centering}
\includegraphics[scale=0.45]{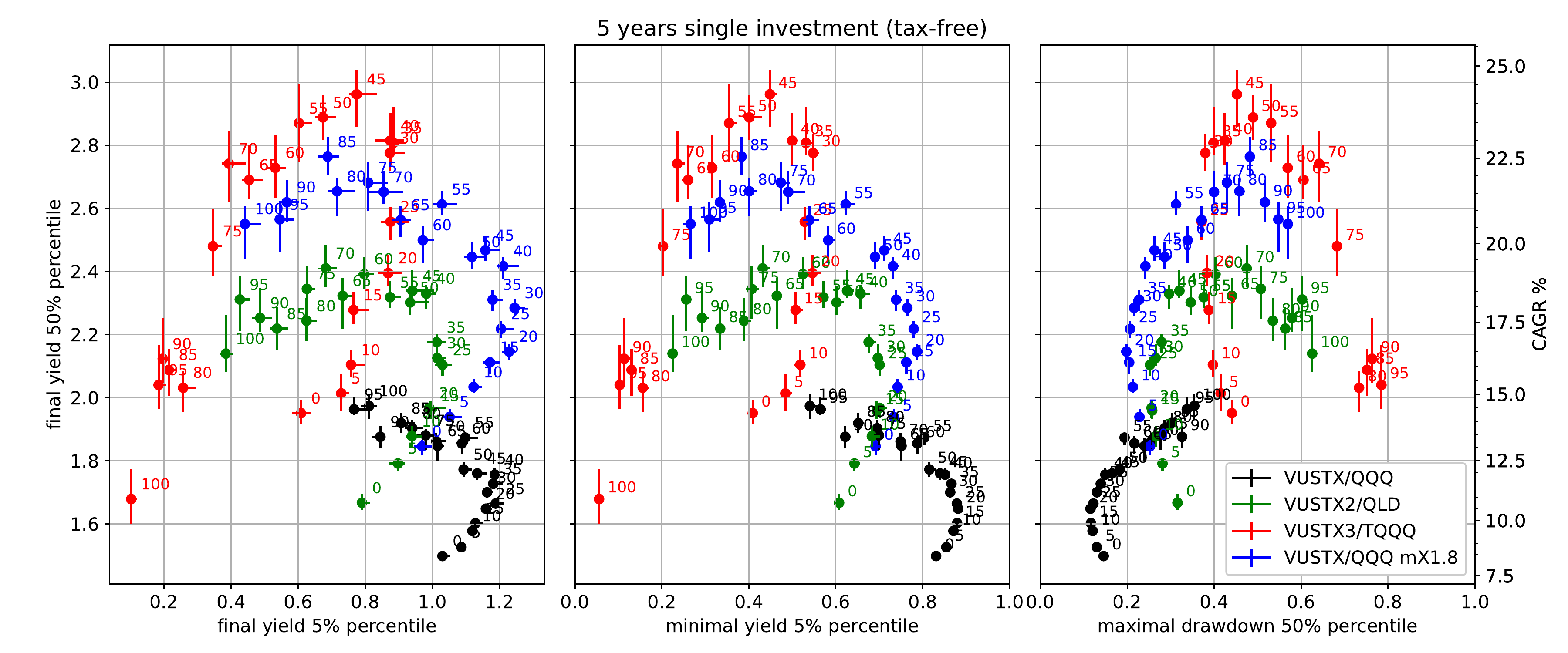}
\par\end{centering}
\caption{NDX100/bonds portfolio metrics for a 5 years tax-free investment.}
\end{figure}
\par\end{center}

We can see that as the investment period is shorter (than 10 years),
the LETF portfolios have a higher (rational) risk. However, a leveraged
mixed portfolio can still beat the unleveraged 100\% stock portfolio
on both risk and reward.

\subsubsection{Tax free (20 years)}
\begin{center}
\begin{figure}[H]
\begin{centering}
\includegraphics[scale=0.45]{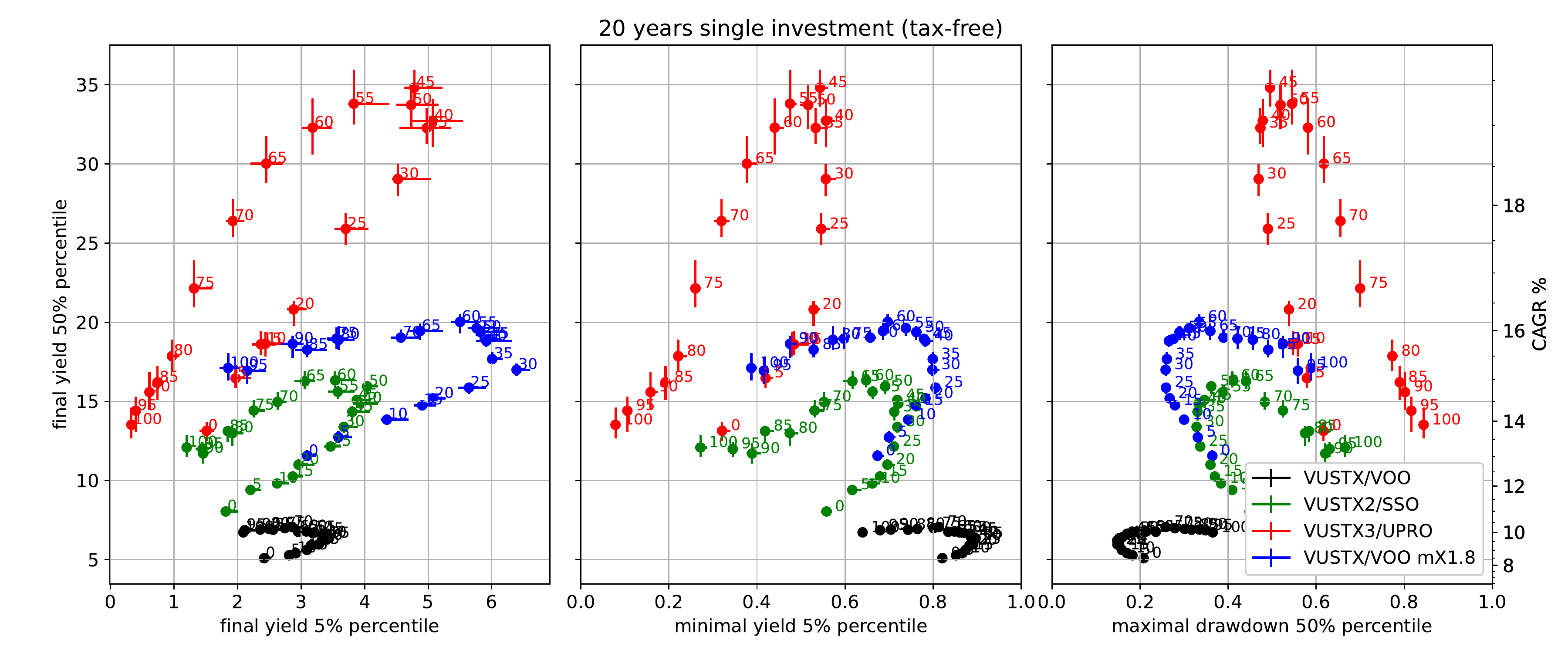}
\par\end{centering}
\caption{SP500/bonds portfolio metrics for a 20 years tax-free investment.}
\end{figure}
\begin{figure}[H]
\begin{centering}
\includegraphics[scale=0.45]{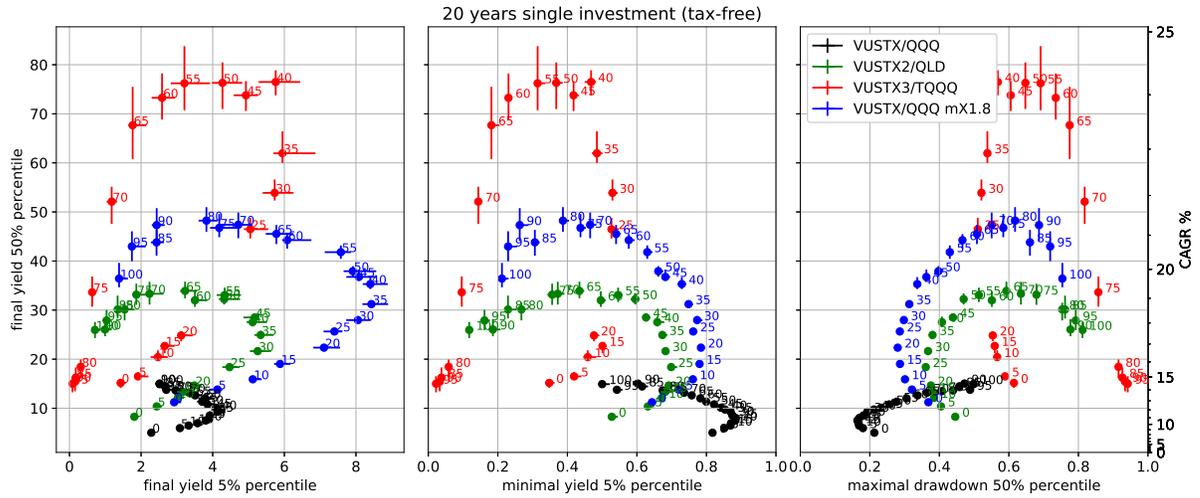}
\par\end{centering}
\caption{NDX100/bonds portfolio metrics for a 20 years tax-free investment.}
\end{figure}
\par\end{center}

In an even longer investment horizon, the opposite happens and the
leveraged portfolios have an even lower (rational) risk and can outperform
even the least risky unleveraged portfolio, on both risk and reward.

Note that the reward for different investments periods (5, 10 and
20 years) is vastly different when looking at the final yield, but
it corresponds to roughly the same CAGR, which is expected.

\subsubsection{Bearish scenario}

The optimistic results we got for the yields are only as good as the
data they were generated from. Meaning, if the daily price change
$\Delta P$ is a probability distribution whose mean is negative,
the investment will not bring profits. The data we used in this work
is from the period 1989-2020. Narrowing the scope around the period
1999-2009 that contains the two major stock crashes of the last 30
years would give much worse results, where stocks do not give positive
yields (bonds still do). 

We define a ``bearish'' scenario using the period 1997-2011. Results
for tax free 10 years investments for the bearish scenario:
\begin{center}
\begin{figure}[H]
\begin{centering}
\includegraphics[scale=0.45]{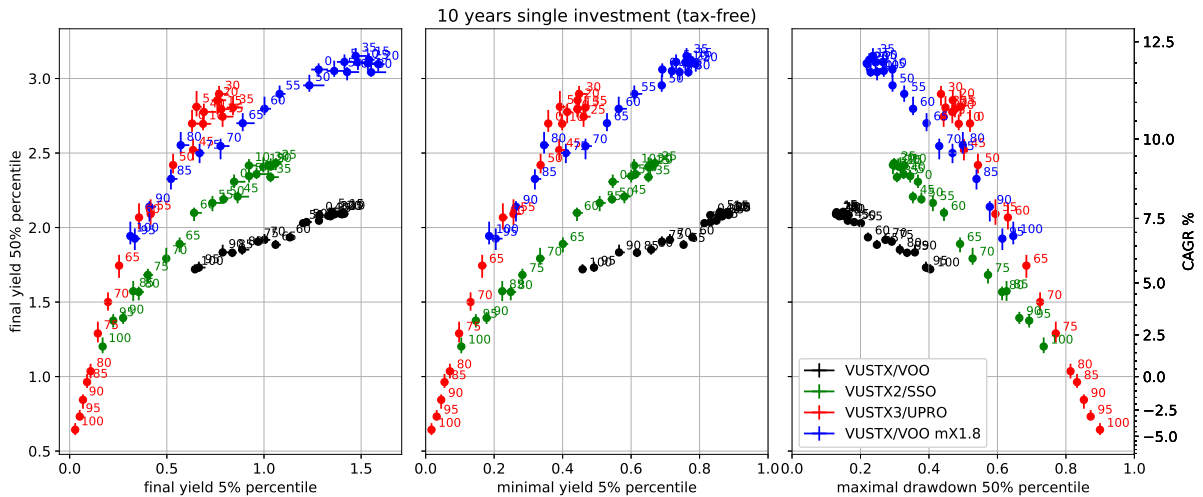}
\par\end{centering}
\caption{SP500/bonds portfolio metrics for a 10 years tax-free investment,
using data from the period 1997-2011 as a ``bearish'' scenario.}
\end{figure}
\par\end{center}

\begin{center}
\begin{figure}[H]
\begin{centering}
\includegraphics[scale=0.45]{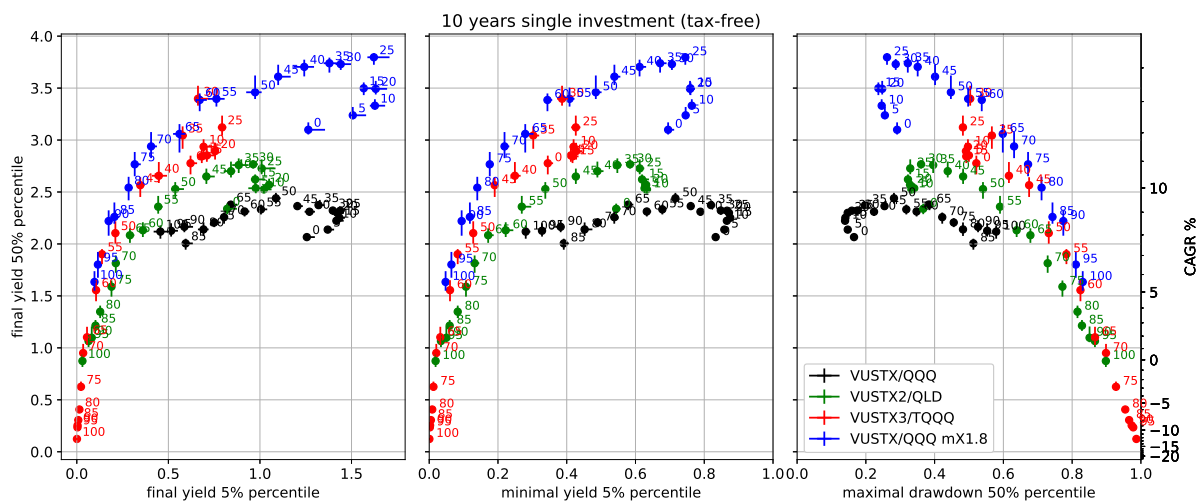}
\par\end{centering}
\caption{NDX100/bonds portfolio metrics for a 10 years tax-free investment,
using data from the period 1997-2011 as a ``bearish'' scenario.}
\end{figure}
\par\end{center}

We can see that in the non-leveraged portfolios, 100\% stocks are
worse than 100\% bonds on both risk and reward, and adding up to $\sim35\%$
stocks is approximately neutral compared to 100\% bonds. This is expected
in the ``bearish'' scenario. Moving to the leveraged LETF portfolios,
we can see higher reward are possible, albeit with higher risk. However,
it can still beat a 100\% stocks portfolio, meaning give higher reward
at the same risk. 

\section{Conclusions}

In this work we performed bootstrapped Monte-Carlo simulations of
leveraged (and unleveraged) mixed portfolios of stocks and bonds,
based on past stock market data (stocks represented by the SP500 and
NDX100 indices, and bonds by the long term treasury ETF VUSTX). We
showed that leverage (using margin debt or leveraged-ETFs) can significantly
amplify the yield of an investment, without significantly increasing
the risk relevant for long-term investors. The rational risk we refer
to is low yield at the end of the investment period. However, the
leveraged portfolios are more volatile and therefore have higher psychological
risks. 

Our results show that for a long term ($>10$ year term) investor
that is happy to take on the risk of 100\% stocks (SP500 or NDX100)
unleveraged portfolio, switching to a 3X leveraged portfolio (using
leveraged-ETFs) with around 35\%-45\% stocks allocation (SP500 or
NDX100) can significantly boost the reward, while reducing the (rational)
risk at the same time. 

If the same investor does not wish to take on more psychological risk
than what the 100\% unleveraged stocks portfolio offers, then switching
to a 2X leveraged portfolio (40\%-60\% SP500 allocation and 35\%-45\%
NDX100 allocation) can deliver higher reward with lower risk (both
rational and psychological). 

We simulate both tax-free and taxable account, and show the conclusions
do not change for a taxable accounts, although the yields are obviously
lower due to the capital gains taxes (25\% tax on profits). In a taxable
account leverage can also be achieved using margin debt, and we show
that a 1.8X margin leveraged portfolio (which can be practically implemented)
slightly outperforms the 2X LETF portfolio, on both risk and reward.

Overall, comparing the two stocks indices, NDX100 outperforms SP500
on both risk and reward.

\end{document}